\documentclass[12pt]{article}

\usepackage{amsfonts}
\usepackage{amsmath}
\usepackage{epsfig}
\usepackage{latexsym}
\usepackage{graphicx}
\usepackage{subfigure}
\usepackage{amssymb}
\numberwithin{equation}{section}
\allowdisplaybreaks

\def\spa#1{\phantom{\fbox{\rule[-#1cm]{0cm}{0cm}}}}

\def\be{\begin{equation}}
\def\ee{\end{equation}}
\def\bea{\begin{eqnarray}}
\def\eea{\end{eqnarray}}
\def\bequ{\begin{equation}}
\def\eequ{\end{equation}}

\def\Tr{\mbox{Tr}}
\def\del{\partial}

\renewcommand{\thefootnote}{\fnsymbol{footnote}}

\def\R{{\mathbb{R}}} 

\newcommand{\eq} {equation}
\newcommand{\eqa} {eqnarray}
\newcommand{\NN} {\mbox {$\nonumber$}}
\newcommand{\eps} {\epsilon}

\newcommand{\bpsi} {\bar{\psi}}
\newcommand{\beps} {\bar{\epsilon}}
\newcommand{\blam} {\bar{\lambda}}

\newcommand{\gam} {\gamma}

\newcommand{\lam} {\lambda}
\newcommand{\vtheta} {\vartheta}
\newcommand{\vphi} {\varphi}
\newcommand{\hQ} {\hat{Q}}
\newcommand{\Det} {\mathrm{Det}}
\newcommand{\vep} {\varepsilon}
\begin{document}

\hfuzz=100pt
\title{{\Large \bf{Higgs branch localization of 3d $\mathcal{N}=2$ theories}} }
\date{}
\author{Masashi Fujitsuka$^a$\footnote{fmasashi@post.kek.jp}, Masazumi Honda$^{b,c,d}$\footnote{masazumihonda@hri.res.in}$\ $  
and Yutaka Yoshida$^{c}$\footnote{yyoshida@post.kek.jp}
  \spa{0.5} \\
\\
$^a${\small{\it Department of Particle and Nuclear Physics,}}
\\ {\small{\it Graduate University for Advanced Studies (SOKENDAI),}}
\\ {\small{\it Tsukuba, Ibaraki 305-0801, Japan}} \\
$^b${\small{\it Harish-Chandra Research Institute,}}
\\ {\small{\it Chhatnag Road, Jhusi, Allahabad 211019, India}} \\
$^c${\small{\it High Energy Accelerator Research Organization (KEK),}}
\\ {\small{\it Tsukuba, Ibaraki 305-0801, Japan}} \\
$^d${\small {\it Yukawa Institute for Theoretical Physics, Kyoto University,}}
\\ {\small {\it Kitashirakawa Oiwakecho, Sakyo-ku, Kyoto 606-8502, Japan}} 
}
\date{\small{December 2013}}

\maketitle
\thispagestyle{empty}
\setcounter{page}{0}
\centerline{}

\begin{abstract}
We study $\mathcal{N}=2$ supersymmetric gauge theories on squashed 3-sphere and $\mathbb{S}^1 \times \mathbb{S}^2$.
Recent studies have shown that 
the partition functions in a class of $\mathcal{N}=2$ theories have factorized forms 
in terms of vortex and anti-vortex partition functions 
by explicitly evaluating matrix integrals obtained 
from
Coulomb branch localization. 
We directly derive this structure by performing Higgs branch localization.
It turns out that more general $\mathcal{N}=2$ theories have this factorization property.
We also discuss the factorization of supersymmetric Wilson loop.
\end{abstract}

\renewcommand{\thefootnote}{\arabic{footnote}}
\setcounter{footnote}{0}

\newpage
\tableofcontents

\section{Introduction}
Recently there has been much progress in understanding non-perturbative aspects of supersymmetric (SUSY) field theories. 
Following to the seminal work by Pestun \cite{Pestun:2007rz}, 
SUSY localization has enabled us to exactly evaluate path integrals in diverse SUSY theories.  
One of the most interesting classes of such theories is three-dimensional $\mathcal{N}= 2$ gauge theory on curved background.
Since the $\mathcal{N}=2$ theories have various dualities expected from string theory
and include low-energy effective theories of M2-branes as special cases,
their better understanding would shed light to non-perturbative understanding of string/M-theory.

In this paper we study the $\mathcal{N}=2$ supersymmetric gauge theories on a squashed 3-sphere and $\mathbb{S}^1 \times \mathbb{S}^2$. 
Several works have shown that 
Coulomb branch localization reduces a class of BPS observables in the $\mathcal{N}= 2$ theories 
to certain matrix integrals \cite{Hama:2011ea,Imamura:2011wg,Drukker:2012sr,Tanaka:2012nr,Closset:2012ru,Kim:2009wb} 
(see also \cite{Kapustin:2009kz,Alday:2013lba,Nian:2013qwa,Nishioka:2013haa}).
Furthermore recent studies \cite{Pasquetti:2011fj,Taki:2013opa} have revealed that 
the partition functions of  some $\mathcal{N}=2$ gauge theories on the squashed 3-sphere 
have the factorization property:
\begin{\eq}
Z \sim \sum_i Z_{\rm V}^{(i)} \bar{Z}_{\rm V}^{(i)} ,
\label{eq:factorization}
\end{\eq}
by explicitly evaluating the matrix integrals.
Here $Z_{\rm V}^{(i)}$ and $\bar{Z}_{\rm V}^{(i)}$ have been expected to be
vortex and anti-vortex functions on $\mathbb{R}^2 \times \mathbb{S}^1$, respectively.
Remarkably, similar structure has also
appeared in 3d 
$\mathcal{N}= 2$ superconfomal indices
in terms of the same (anti-)vortex partition function \cite{Dimofte:2011py,Hwang:2012jh}.
This implies that
partition functions of 3d $\mathcal{N}= 2$ theories on various spaces 
consist of the same building blocks
referred to as holomorphic block \cite{Beem:2012mb} (see also \cite{Nieri:2013yra,Nieri:2013vba,Imamura:2013qxa}).
Thus one can expect that the holomorphic block is more fundamental quantity in the 3d $\mathcal{N}= 2$ theories.
Despite of such illuminating structure, we still lack for direct understanding of this structure.
In this paper 
we directly derive the vortex structure by performing Higgs branch localization as in two dimensions \cite{Benini:2012ui,Doroud:2012xw},
and give precise identifications of parameters in the vortex partition functions.
Moreover we find that the factorization property appears in more general $\mathcal{N}=2$ theories 
than the ones studied in \cite{Pasquetti:2011fj,Taki:2013opa}.
We also discuss the factorization property of supersymmetric Wilson loop.

This paper is organized as follows.
We briefly introduce the $\mathcal{N}=2$ supersymmetric gauge theory on 3d ellipsoid in section~\ref{sec:construction},
which is review of \cite{Hama:2011ea,Drukker:2012sr}.
We also review Coulomb branch localization by using index theorem \cite{Drukker:2012sr} in section~\ref{sec:Coulomb}
because
this computation is quite useful also for Higgs branch localization.
In section~4, we perform Higgs branch localization and
show that part of saddle points in path integral are given by BPS vortex configurations.
In section~5, we introduce vortex world line theory for $\mathcal{N}=2$ supersymmetric gauge theory and evaluate the
vortex partition function via localization. 
In other words,
we explicitly show that
the function $Z_{\rm V}^{(i)}$($\bar{Z}_{\rm V}^{(i)}$) appeared in \eqref{eq:factorization}
are actually the (anti-)vortex function on $\mathbb{R}^2 \times \mathbb{S}^1$ in our situation.
In section~6, we discuss the factorization property of BPS Wilson loop on the squashed sphere.
In section~7, we provide some interesting examples.
In section~8, we apply Higgs branch localization to superconformal index.
Section~9 is devoted to conclusion.

\subsubsection*{Note added}
When our paper was ready\footnote{
Preliminary version of our results were presented by M.~F. 
at ``The Physical Society of Japan Autumn Meeting 2013'', Kochi University, Japan September 20-23 2013.
} for submission to the arXiv,
there appeared a paper \cite{Chen:2013pha} which has a few overlaps with ours.

\section{$\mathcal{N}=2$ supersymmetric gauge theory on the 3d ellipsoid}
\label{sec:construction}
In this section we briefly introduce $\mathcal{N}=2$ supersymmetric gauge theory on three dimensional ellipsoid.
This section is essentially review of \cite{Hama:2011ea,Drukker:2012sr}.
\subsection{Three dimensional ellipsoid}
In this paper we choose 3d ellipsoid $\mathbb{S}_{b}^{3}$ as a squashed 3-sphere, which is defined by the following hypersurface\footnote{
There are also other choices of squashed spheres \cite{Imamura:2011wg,Alday:2013lba,Nishioka:2013haa} (see also \cite{Nian:2013qwa,Martelli:2013aqa}).
However they 
present the same partition function  
as long as we consider 1-parameter deformation of the round sphere with keeping SUSY \cite{Closset:2013vra}. 
}:
\begin{\eq}
x_0^2 +x_1^2 +x_2^2 +x_3^2 =1 ,
\end{\eq}
in the space with the metric
\begin{\eqa}
ds^2 = l^2 (dx_0^2 +dx_1^2 ) +\tilde{l}^2 (dx_2^2 +dx_3^2 ) .
\end{\eqa}
There are two convenient coordinates in our context. 
First one is torus fibration coordinate:
\begin{\eqa}
&&(x_0 , x_1 , x_2 , x_3 )
= ( \cos{\vtheta}\cos{\vphi_2} , \cos{\vtheta}\sin{\vphi_2} , \sin{\vtheta}\cos{\vphi_1} , \sin{\vtheta}\sin{\vphi_1} ) ,
\end{\eqa}
with $0\leq \vtheta \leq \pi /2 , 0\leq \vphi_1 < 2\pi , 0\leq \vphi_2 < 2\pi $.
In this coordinate, the metric and  orthogonal frame are given by
\begin{\eqa}
&&ds^2 
= R^2 \left( f(\vtheta )^2 d\vtheta^2 +b^2 \sin^2{\vtheta}d\vphi_1^2 +b^{-2}\cos^2{\vtheta}d\vphi_2^2 \right) ,\NN\\
&& e^1 = Rb^{-1}\cos{\vtheta}d\vphi_2 ,\quad e^2 = -Rb \sin{\vtheta}d\vphi_1 ,\quad e^3 = Rf(\vtheta )d\vtheta ,
\end{\eqa}
where
\begin{\eq}
R=\sqrt{l\tilde{l}},\quad  b=\sqrt{\tilde{l}/l},\quad
f(\vtheta ) = \sqrt{b^{-2}\sin^2{\vtheta} +b^2 \cos^2{\vtheta}} .
\end{\eq} 
Killing spinors in this space satisfy
\begin{\eq}
D_\mu \eps  =\frac{i}{2Rf(\vtheta )} \gam_\mu  \eps ,\quad
D_\mu \beps =\frac{i}{2Rf(\vtheta )} \gam_\mu  \beps , \label{KS-eq}
\end{\eq}
where the covariant derivative is defined by turning on a background $U(1)$ gauge field $V=\frac12(1-\frac{b}{f})d\varphi_1 +\frac12 (1-\frac{b^{-1}}{f})d\varphi_2$ additionally.
These equations are solved by \cite{Hama:2011ea}\footnote{
Our notation is slightly different from \cite{Hama:2011ea}: $(\vtheta,\varphi_1,\varphi_2)_{\text{here}}=(\theta,-\chi,\varphi)_{\text{there}}$, $(\eps, \beps)_{\text{here}}=(-\beps,\eps)_{\text{there}}$.}
\begin{\eq}
\eps = \frac{1}{\sqrt{2}} 
       \begin{pmatrix} e^{\frac{i}{2}(\vphi_1 +\vphi_2 +\theta )} \cr 
                       e^{\frac{i}{2}(\vphi_1 +\vphi_2 -\theta  )} \end{pmatrix},\quad
\beps = \frac{1}{\sqrt{2}} 
       \begin{pmatrix} -e^{-\frac{i}{2}(\vphi_1 +\vphi_2 -\theta )} \cr 
                        e^{-\frac{i}{2}(\vphi_1 +\vphi_2 +\theta )} \end{pmatrix} .
\end{\eq}
The other convenient coordinate is Hopf fibration coordinate given by
\begin{\eq}
\vtheta =\frac{1}{2} \theta ,\quad \vphi_1 =\frac{1}{2}(\psi -\phi ),\quad \vphi_2 =\frac{1}{2}(\psi +\phi ) .
\end{\eq}
Note that its ``$\mathbb{S}^2$ part'' has north and south poles at $\theta =0$ and $\theta =\pi$, respectively.

\subsection{Vector multiplet}
\label{sec:vec_action}
Let us start with $\mathcal{N}=2$ vector multiplet on $\mathbb{S}_b^3$.
The action of $\mathcal{N}=2$ super Yang-Mills theory (SYM) is given by
\begin{\eqa}
S_{\text{YM}}
&=&  \frac{1}{g_{\text{YM}}^2}\int d^3 x \sqrt{g}\,\Tr \Big{[}
     \frac14 F_{\mu\nu}F^{\mu\nu}+\frac12 D_\mu \sigma D^\mu \sigma +\frac12 \Big{(} D+\frac{\sigma}{R f(\vtheta)} \Big{)}^2 \NN \\
&&   ~~~~~~~~~~~~~~~~~~~~~~~+\frac{i}{2}\blam \gamma^\mu D_\mu \lambda +\frac{i}{2}\blam [\sigma, \lambda] -\frac{1}{4R f(\vtheta)}\blam \lambda \Big{]} .
\label{SYMaction}
\end{\eqa}
This action is invariant under the following SUSY transformation\footnote{
See appendix.\ref{app:SUSY} for detail.
 }:
\begin{\eq}
\begin{split}
Q A_\mu   &= -\frac{1}{2} \blam \gam_\mu \eps -\frac{1}{2} \beps \gamma_\mu \lambda, \\
Q \sigma  &= \frac{i}{2} \blam\eps +\frac{i}{2} \beps \lambda, \\
Q \lambda &= \left( \frac{1}{2} \eps_{\mu\nu\rho}F^{\nu\rho} - D_\mu \sigma \right) \gam^\mu \eps -i D\eps  -\frac{i}{Rf(\vtheta )} \sigma \eps ,\\
Q \blam   &= \left( \frac{1}{2} \eps_{\mu\nu\rho}F^{\nu\rho} + D_\mu \sigma  \right) \gam^\mu \beps +iD\beps +\frac{i}{Rf(\vtheta )} \sigma \beps ,\\
Q D       &= \frac{1}{2}\beps \gam^\mu D_\mu \lambda 
                   -\frac{1}{2}D_\mu \blam \gam^\mu \eps 
                   -\frac{1}{2}[\beps \lambda ,\sigma ]
                   +\frac{1}{2}[\blam\eps ,\sigma ] 
                   +\frac{1}{4Rf(\vtheta)}(\beps \lambda -\blam \eps ),  \label{Q-vec}
\end{split}
\end{\eq}
with the ``commuting`` spinors $\eps$ and $\beps$.
One can show that $Q^2$ generates 
\begin{\eq}
Q^2 = i\mathcal{L}_v +i\sigma -v^\mu A_\mu +\frac{1}{2R}(b+b^{-1})\mathcal{R} ,
\label{Qsq}
\end{\eq}
where
\begin{\eq}
v = \beps \gam^{\hat{\mu}} \eps e_{\hat{\mu}} 
=R^{-1} \left( b^{-1}\frac{\del}{\del\varphi_1} +b\frac{\del}{\del\varphi_2} \right) ,
\end{\eq}
$\mathcal{L}_v$ is a Lie-derivative along the $v$ and $\mathcal{R}$ is the R-symmetry generator. 
Note that we can write the SYM Lagrangian to be $Q$-exact up to total derivatives:
\begin{\eq}
\mathcal{L}_{\text{YM}}
=Q V_{\text{vec}} ,\quad {\rm with}\ 
V_{\text{vec}} =  {\rm Tr} \frac{(Q\lam )^\dag \lam +(Q \blam )^\dag \blam }{4}  , 
\label{def-YM}
\end{\eq}
if we take the integral contour of $D$ to be real. 
We can also consider supersymmetric Chern-Simons term and Fayet-Illiopoulos (FI) term as
\begin{\eqa}
&&S_{\rm CS} = \frac{i\kappa}{4\pi}\int d^3 x \sqrt{g} {\rm Tr} 
                  \Biggl[ \eps^{\mu\nu\rho} \left( A_\mu \del_\nu A_\rho +\frac{2i}{3}A_\mu A_\nu A_\rho \right)
                            -\blam\lambda +2D\sigma  \Biggr] ,\\
&&S_{\rm FI} = -\frac{i\zeta}{2\pi R} \int d^3 x \sqrt{g}\left( D -\frac{1}{Rf(\vtheta )} \sigma \right) ,
\end{\eqa}
respectively.

\subsubsection*{BPS configuration}
From the $Q$-exact SYM action \eqref{SYMaction},
we immediately find the following BPS configuration
\begin{\eqa}
F_{\mu\nu} =0 ,\quad D_\mu \sigma =0 ,\quad D =-\frac{1}{Rf(\vtheta )} \sigma .
\label{naiveBPS}
\end{\eqa}
This is nothing but the saddle point used for the Coulomb branch localization \cite{Hama:2011ea,Drukker:2012sr}.
Note that this configuration does not include BPS vortex configurations,
which we desire in Higgs branch localization.
One can show that 
relaxing the reality condition of $D$ in \eqref{def-YM}
enables us to find the wider BPS configuration 
\begin{\eq}
\begin{split}
&F_{12}=0.\quad
F_{23}+\text{Im}D\cos \vartheta =0,\quad
F_{31}-\text{Im}D\sin \vartheta =0, \\
&D_\mu \sigma=0,\quad
\text{Re}D = -\frac{\sigma}{Rf(\vartheta)} , \label{BPS-eq}
\end{split}
\end{\eq}
which 
does not contradict
the BPS vortex configuration\footnote{
The same procedure has been performed in the 2-dimensional case \cite{Benini:2012ui}.}.
As we will see later,
an appropriate choice of a deformation term 
leads us to a natural change of the integral contour of $D$ from real to complex and
serves a nontrivial $\text{Im}D$.
Then we will obtain the desired vortex configurations at the north pole ($\vtheta =0$) and south pole ($\vtheta =\pi /2$)
as the saddle points of localization. 
\subsection{Chiral multiplet}
Let us consider matter sector. The action is given by
\begin{\eqa}
S_{\text{mat}}
&=&  \int d^3x \sqrt{g}\,\Big{(}
          D_\mu \bar{\phi}D^\mu \phi +\bar{\phi}\sigma^2 \phi +\frac{i(2\Delta-1)}{Rf(\vtheta)}\bar{\phi}\sigma\phi
          +\frac{\Delta(2-\Delta)}{(R f(\vtheta))^2}\bar{\phi}\phi +i \bar{\phi}D \phi +\bar{F}F \NN \\
&&~~~~~~~~~~~~~~-i \bpsi \gamma^\mu D_\mu \psi +i \bpsi \sigma\psi -\frac{2\Delta-1}{2R f(\vtheta)}\bpsi \psi 
          +i\bpsi \lambda \phi -i \bar{\phi}\blam \psi  \Big{)} , \label{matlag}
\end{\eqa}
which is invariant under the SUSY transformation
\begin{\eq}
\begin{split}
Q \phi      &= i\beps \psi ,\\
Q \bar{\phi}&= i\eps \bpsi ,\\
Q \psi      &= - \gam^\mu \eps D_\mu \phi -\eps \sigma \phi
                       -\frac{i\Delta}{Rf(\vtheta )} \eps \phi +i\beps F ,\\
Q \bpsi     &= -\gam^\mu \beps D_\mu \bar{\phi} -\bar{\phi} \sigma \beps
                        -\frac{i\Delta}{Rf(\vtheta )} \bar{\phi} \beps   +i\bar{F}\eps , \\
Q F         &= \eps (-\gam^\mu D_\mu \psi +\sigma \psi +\lambda \phi ) 
                     +\frac{i}{2Rf(\vtheta )}(2\Delta -1)  \eps \psi ,\\
Q \bar{F}    &= \beps (-\gam^\mu D_\mu \bar{\psi} +\bar{\psi}\sigma  -\bar{\phi}\blam ) 
                     +\frac{i}{2Rf(\vtheta )}(2\Delta -1)  \beps \bar{\psi}  .  \label{Q-mat}
\end{split}
\end{\eq}
We have assigned R-charges:$(-\Delta ,\Delta , 1-\Delta ,\Delta -1 ,2-\Delta ,\Delta -2)$
to $(\phi ,\bar{\phi}, \psi ,\bar{\psi} ,F ,\bar{F})$, respectively.
For the Coulomb branch localization, the authors in \cite{Drukker:2012sr} have used the following deformation term:
\begin{\eq}
\mathcal{L}_\psi  = Q V_{\text{chi}} ,\quad {\rm with}\ 
V_{\text{chi}} =   \frac{(Q\psi )^\dag \psi +(Q\bpsi )^\dag \bpsi}{2} . 
\label{def-mat}
\end{\eq}
Completing square leads us to 
\begin{\eqa}
\left. \mathcal{L}_\psi \right|_{\rm Bos.}
&=&  \left| \sin{\vtheta}D_1 \phi +\cos{\vtheta}D_2 \phi +iD_3 \phi \right|^2 
              +|\sigma \phi |^2  +|F|^2   \NN\\
&&              +\left| \cos{\vtheta}D_1 \phi -\sin{\vtheta}D_2 \phi +\frac{i\Delta}{Rf(\vtheta )}\phi \right|^2 .
\label{mat-lag}
\end{\eqa}

\subsubsection*{BPS configuration}
From (\ref{mat-lag}), we find BPS configuration for chiral multiplet as
\begin{\eqa}
&&\sin{\vtheta}D_1 \phi +\cos{\vtheta}D_2 \phi +iD_3 \phi =0,\quad
\sigma \phi =0,\quad F=0 ,\NN\\
&&\cos{\vtheta}D_1 \phi -\sin{\vtheta}D_2 \phi +\frac{i\Delta}{Rf(\vtheta )}\phi =0 .
\end{\eqa}

\section{Localization on Coulomb branch}
\label{sec:Coulomb}
In this section we review the Coulomb branch localization of the $\mathcal{N}=2$ theories on the ellipsoid by the index theorem.
Although this has been already done in \cite{Drukker:2012sr},
the computation technique is quite useful also for Higgs branch localization as we will see in the next section.
Choosing the deformation term $QV$ as
\begin{\eq}
QV = \mathcal{L}_{\text{YM}}+\mathcal{L}_{\psi} ,
\end{\eq}
we find the following Coulomb branch localized configuration:
\begin{\eq}
A_\mu =0 ,\quad \sigma = \text{const.} ,\quad D =-\frac{1}{Rf(\vtheta )} \sigma ,\quad  \phi=F=0. 
\label{Coul}
\end{\eq}
Our remaining task
is only to compute the one-loop determinant around the saddle point. 
\subsection{Gauge fixing}
In order to compute the one-loop determinant, we have to perform gauge fixing.
We introduce BRST transformation by
\begin{\eqa}
 Q_B A_\mu = D_\mu c ,\quad  Q_B c       = -\frac{i}{2}[c,c], \quad  Q_B \bar{c} = B, \quad  Q_B B=0 ,
\end{\eqa}
where $c$, $\bar{c}$ are ghosts and $B$ is Nakanishi-Lautrap field. 
Then we find gauge fixing action as 
\begin{\eq}
\mathcal{L}_{\rm gh} =Q_B V_{\rm gh} = Q_B {\rm Tr}\Biggl[ \bar{c} \left( G(\tilde{A}) +\frac{\xi}{2}B \right) \Biggr] ,
\end{\eq}
where $G(\Phi )$ is the gauge fixing function and $\tilde{\Phi}$ stands for the fluctuation from the localized configuration $\Phi^{(0)}$ given by
\begin{\eq}
\tilde{\Phi} =\Phi -\Phi^{(0)} .
\end{\eq}
We also specify SUSY transformations for ghosts and Nakanishi-Lautrap field as
\begin{\eqa}
 Q c = \tilde{\sigma} +iv^\mu \tilde{A}_\mu ,\quad
 Q\bar{c} =0 ,\quad  Q B = iv^\mu D_\mu^{(0)}\bar{c} +i[\sigma^{(0)} ,\bar{c} ] .
\end{\eqa}
Then 
$\hat{Q} = Q+Q_B$ 
generates 
\begin{\eq}
\hat{Q}^2 = i\mathcal{L}_v +i\sigma^{(0)} -v^\mu A_\mu^{(0)} +\frac{1}{2R}(b+b^{-1}) \mathcal{R} . \label{hatQ^2}
\end{\eq}
Although we had omitted the gauge fixing in the above sections for simplicity,
precisely speaking, we have to choose the deformation term as
\begin{\eq}
QV \quad  \to  \quad  \hat{Q}\hat{V},\quad \text{where}\quad \hat{V} =V+V_{\rm gh} .
\end{\eq}

\subsection{One-loop determinant}
We compute the one-loop determinant around \eqref{Coul}
by the index theorem along the argument in \cite{Drukker:2012sr}. 
First it is convenient to define bosonic and fermionic coordinates $(X_0, X_1)$ as, 
\begin{\eqa}
&&X_0 = (X_0^{\rm vec} ;X_0^{\rm chi } )=(\tilde{A}_\mu ;\phi ,\bar{\phi} ),\quad
  X_1 = (X_1^{\rm vec} ;X_1^{\rm chi } )=(\Lambda ,c,\bar{c} ;  \eps\psi ,\beps\bpsi ) ,
\end{\eqa}
where
\begin{\eq}
\Lambda = \beps \lambda +\eps\blam .
\end{\eq}
Then we can write quadratic fluctuation part of $\hQ \hat{V}$ as
\begin{\eqa}
\hQ \hat{V} |_{\rm quad}
&=& ( X_0 ,\hQ X_1 ) 
  \begin{pmatrix} \hQ^2 & 0 \cr 0 & 1 \end{pmatrix}  
  \begin{pmatrix} D_{00} & D_{01} \cr D_{10} & D_{11} \end{pmatrix}  
  \begin{pmatrix} X_0 ,\hQ X_1 \end{pmatrix}  \NN\\
&& -(\hQ X_0 ,X_1 ) \begin{pmatrix} D_{00} & D_{01} \cr D_{10} & D_{11} \end{pmatrix}  
  \begin{pmatrix} 1 & 0 \cr 0 & \hQ^2 \end{pmatrix}  
  \begin{pmatrix} \hQ X_0 , X_1 \end{pmatrix} .
\end{\eqa}
Since $[\hQ^2 ,D_{10}]=0$ and non-zero modes of $D_{10}$ are paired,
the one-loop determinant is simply given by
\begin{\eq}
Z_{\rm 1-loop} = \left( \frac{\det{\hQ}^2 |_{{\rm CoKer}D_{10}}}{\det{\hQ^2}|_{{\rm Ker}D_{10}}} \right)^{\frac{1}{2}} .
\end{\eq}
Therefore once we know the equivariant index ${\rm ind}D_{10}$, 
we can find the one-loop determinant 
by the rule
\begin{\eq}
{\rm ind}D_{10} = \sum_j c_j e^{w_j} \quad \rightarrow \quad Z_{\rm 1-loop} = \prod_j w_j^{-\frac{c_j}{2}} . \label{ind-rule}
\end{\eq}
Note that although these operators are infinite dimensional, 
the index is well-defined when $D_{10}$ is at least transversally elliptic \cite{Atiyah}.
Thus our problem is reduced to compute the equivariant index. 
Although we would like to use the index theorem to obtain the index as 
in
\cite{Pestun:2007rz, Benini:2012ui}, there is no fixed point on $\mathbb{S}_{b}^{3}$.
The authors of \cite{Drukker:2012sr} have resolved this problem in the following way.
First we rewrite the vector field $v$ 
in terms of the Hopf-fibration coordinate as  
\begin{\eqa}
v=\bar{\epsilon}\gamma^\mu \epsilon\, \del_\mu 
=  \frac{1}{R}\Big{(} b^{-1}\del_{\varphi_1} +b\, \del_{\varphi_2} \Big{)} 
=  \frac{1}{R}\Big{(} (b+b^{-1})\del_\psi +(b-b^{-1})\del_{\phi} \Big{)}.
\label{Liederiv}
\end{\eqa}
Here we have two $U(1)$ actions generated by $\del_{\psi}$ and $\del_{\phi}$. 
Especially $\del_{\psi}$ rotates the Hopf fiber and acts on $\mathbb{S}_{b}^3$ freely. 
In fact we can show that each $D_{10}$ in vector and chiral multiplets is transversally elliptic with respect to these actions. 
It is known that
when a part of the group action is free, a transversally elliptic operator can be reduced to the one on the quotient space \cite{Atiyah}.
Namely, $D_{10}$ 
is reduced to the one 
on the base $\mathbb{S}^{2}$  in our case. 
Then the index theorem says that we have only to compute contributions from fixed points of $\del_{\phi}$-action, 
which are the north pole ($\theta =0$) and south pole ($\theta =\pi$). 
As a result, the authors in \cite{Drukker:2012sr} have found the indices as
\begin{\eqa}
\text{ind}(D_{10}^{\text{chi}})
&=& 2\Big{(} \exp\Big{[}\prod_{\omega \in R}\prod_{m=0}^{\infty}\prod_{n=0}^{\infty}\,
             i\Big{\{}mb+nb^{-1}+\frac{Q}{2}-i\omega(\hat{\sigma})-\frac{Q}{2}(1-\Delta) \Big{\}} \Big{]}\NN \\
&&     - \exp\Big{[}\prod_{\omega \in R}\prod_{m=0}^{\infty}\prod_{n=0}^{\infty}\,
             -i\Big{\{}mb+nb^{-1}+\frac{Q}{2}+i\omega(\hat{\sigma})+\frac{Q}{2}(1-\Delta)\Big{\}} \Big{]} \Big{)} ,\NN\\
\text{ind}(D_{10}^{\text{vec}}) 
&=& -\sum_{n\in\mathbb{Z}}e^{inb} \sum_{\alpha}e^{\alpha(\hat{\sigma})}
    -\sum_{n\in\mathbb{Z}}e^{inb^{-1}} \sum_{\alpha}e^{\alpha(\hat{\sigma})},
\end{\eqa}
where $Q \equiv b+b^{-1}$, $\hat{\sigma}\equiv R \sigma$, $\omega$ and $\alpha$ denote weights in representation $R$ 
and the roots in the gauge group. 
Thus applying the rule (\ref{ind-rule}) leads us to the one-loop determinants 
\begin{\eqa}
&&Z_{\rm chi}^{\rm (1-loop )}=\prod_{w\in R} s_b \left( \frac{iQ}{2}(1-\Delta ) -w(\hat{\sigma})\right) ,\NN\\
&&Z_{\rm vec}^{\rm (1-loop )}=\prod_{\alpha > 0} \sinh (\pi b \alpha (\hat{\sigma})) \sinh (\pi b^{-1}\alpha (\hat{\sigma})) .
\end{\eqa}

\section{Localization on Higgs branch}
\label{sec:Higgs_loc}
In this section we perform Higgs branch localization of the ellipsoid partition function.

\subsection{Localized configuration}
As a deformation term, we take\footnote{
Below we take $R=1$.
Dependence on $R$ can be easily recovered by dimensional analysis.
}
\begin{\eq}
\mathcal{L}_{\rm YM} +\mathcal{L}_{\psi}  +\mathcal{L}_{ H} . 
\end{\eq}
While we have used 
$\mathcal{L}_{\rm YM} $ \eqref{def-YM} and $\mathcal{L}_{\psi}$ \eqref{def-mat}
in the Coulomb branch localization,
$\mathcal{L}_H $ is the new deformation term for our Higgs branch localization defined by
\begin{\eq}
\mathcal{L}_H =Q V_H
= i^{-1}Q{\rm Tr}\Bigl[ \frac{(\eps^\dag \lambda -\beps^\dag \blam )h}{4i} \Bigr] ,
\label{LH}
\end{\eq}
as in the two dimensional case \cite{Benini:2012ui,Doroud:2012xw}.
Here $h$ is a function of scalars and we take $h$ case by case depending on the field content.
For example, if we consider $\mathcal{N}=2$ theory with fundamental and anti-fundamental chiral multiplets 
including scalar fields $\phi$ and $\tilde{\phi}$, respectively,
then we choose $h$ as 
\begin{\eq}
h=\phi \phi^\dag -\tilde{\phi}^\dag \tilde{\phi} -\chi \mathbf{1}_N ,
\end{\eq}
where $\chi$ is a parameter\footnote{
A physical meaning of $\chi$ is
FI-parameter in vortex world line theory as we will see below.
The limit $\chi  \rightarrow \pm\infty$ corresponds to
the limit in which the vortex becomes point-like.
} taken to be $\chi  \rightarrow \pm\infty$ later.
When we further add an adjoint chiral multiplet with its scalar $X$, $h$ becomes
\begin{\eq}
h=[X,X^\dag ] +\phi \phi^\dag -\tilde{\phi}^\dag \tilde{\phi} -\chi \mathbf{1}_N .
\end{\eq}

In terms of $h$, we can write the Bosonic part of $\mathcal{L}_H$ as
\begin{\eqa}
 \left. \mathcal{L}_H \right|_{\rm Bos.}
&=&  {\rm Tr}\Biggl[ \left( -\frac{1}{2}\cos{\vtheta} F_{23}   +\frac{1}{2}\sin{\vtheta} F_{31} 
                          +\frac{i}{2}D +\frac{i}{2f(\vtheta )} \sigma  \right) h \Biggr] .
\end{\eqa}
Combined with $\mathcal{L}_{\rm YM}$,
completing square leads us to
\begin{\eqa}
&&\left. \mathcal{L}_{\rm YM} \right|_{\rm Bos.} +\left. \mathcal{L}_H \right|_{\rm Bos.} \NN\\
&&=  {\rm Tr}\Biggl[  \frac{1}{2}F_{12}^2 
                    +\frac{1}{2} \left( \sin{\vtheta} F_{23} +\cos{\vtheta}F_{31} \right)^2
                    +\frac{1}{2} \left( \cos{\vtheta} F_{23} -\sin{\vtheta}F_{31} -\frac{1}{2}h \right)^2 \NN\\
&&~~~~~~~   +\frac{1}{2}(D_\mu \sigma )^2
     +\frac{1}{2}\left( D + \frac{1}{f(\vtheta )}\sigma +\frac{i}{2}h \right)^2  \Biggr] .
\label{eq:D_int}
\end{\eqa}
Note that $D$ can be trivially integrated out and
this becomes semi-positive definite after that. 
Since $\mathcal{L}_\psi$ is also semi-positive definite itself,
we obtain the following localized configuration:
\begin{\eqa}
&& F_{12}=0,\quad \sin{\vtheta} F_{23} +\cos{\vtheta}F_{31} =0,\quad
  \cos{\vtheta} F_{23} -\sin{\vtheta}F_{31} -\frac{1}{2}h =0 ,\NN\\
&& D_\mu \sigma =0,\quad D + \frac{1}{f(\vtheta )}\sigma +\frac{i}{2}h =0, \quad
 \sin{\vtheta}D_1 \phi +\cos{\vtheta}D_2 \phi +iD_3 \phi =0,\NN\\
&&   \cos{\vtheta}D_1 \phi -\sin{\vtheta}D_2 \phi +\frac{i\Delta}{f(\vtheta )}\phi =0,\quad
  \sigma \phi =0,\quad F=0 .
\label{eq:localizedH0}
\end{\eqa}

\subsubsection*{Remark}
Originally we have taken the integral contour of $D$ to be real.
For integrating $D$ out,
we have to change the integral contour of $D$ from $\mathbb{R}$ to $\mathbb{R}-ih/2$
as seen from \eqref{eq:D_int}. 
As we briefly mentioned in the last of sec.~\ref{sec:vec_action},
this gives the imaginary part of $D$ and
hence we obtain the BPS configuration of the type \eqref{BPS-eq}
consistent with vortex configurations.

\subsection{Away from north and south poles}
First let us specify our theory to $\mathcal{N}=2\ U(N)$ gauge theory 
with $N_f$ fundamental chiral multiplets as a warm up.
We also take $\Delta =0$ for simplicity, and introduce a real mass $M$.
Note that since $Q^2$ is holomorphic with respect to $M +i(b+b^{-1})\frac{\Delta}{2}$,
we can reproduce the result for general $\Delta$ from the one for $\Delta =0$ by the analytic continuation of $M$.
For this case, the localized configuration is
\begin{\eqa}
&& F_{12}=0,\quad \sin{\vtheta} F_{23} +\cos{\vtheta}F_{31} =0,\quad
  \cos{\vtheta} F_{23} -\sin{\vtheta}F_{31} -\frac{1}{2}(\phi \phi^\dag -\chi \mathbf{1}_N) =0 ,\NN\\
&& D_\mu \sigma =0,\quad D + \frac{1}{f(\vtheta )}(\sigma +M)  +\frac{i}{2}(\phi \phi^\dag -\chi \mathbf{1}_N) =0, \NN\\
&& \sin{\vtheta}D_1 \phi +\cos{\vtheta}D_2 \phi +iD_3 \phi =0,\quad
   \cos{\vtheta}D_1 \phi -\sin{\vtheta}D_2 \phi   =0,\NN\\
&&  (\sigma +M) \phi =0,\quad F=0 .
\label{eq:localizedH}
\end{\eqa}
Away from the north and south poles, we consider only smooth solutions.
Imposing the smoothness condition, we easily find that $\phi$ should be constant.
Although $\phi$ can be arbitrary constant values as long as $(\sigma +M) \phi =0$ is satisfied,
we can show that only ones such that $\phi\phi^\dag =\chi \mathbf{1}_N$ has a nonzero contribution in the limit $\chi\rightarrow\pm \infty$ as follows.
From eq.~\eqref{eq:localizedH}
we explicitly write the field strength as
\begin{\eqa}
F_{\mu\nu}dx^\mu \wedge  dx^\nu 
&=& d\Biggl[ \frac{ (\phi \phi^\dag -\chi )}{6(b^2 -b^{-2})}  f^3 (\vtheta ) (bd\vphi_1 -b^{-1}d\vphi_2 ) \Biggr]   .
\end{\eqa}
Since the gauge field then satisfies $v^\mu A_\mu =0$ up to gauge choice,
the one-loop determinant should be $\chi$-independent and 
therefore $\chi$-dependence appears only on the classical contribution.
If we have FI-term or Chern-Simons term, which is our case, 
then this gives exponentially suppression factor $\sim e^{-|\chi |}$  and 
vanishes in the $|\chi |\rightarrow \infty$ limit\footnote{
Strictly speaking, we take the limit $\chi \rightarrow -{\rm sgn}(\zeta +\kappa\sum_i m_{l_i} )\infty$.
} except for $\phi\phi^\dag =\chi \mathbf{1}_N$.

Thus we conclude that non-vanishing smooth configuration is only the Higgs branch solution
\begin{\eq}
F_{\mu\nu} =0,\quad D_\mu \sigma =0,\quad (\sigma +M)\phi =0,\quad \phi \phi^\dag -\chi\mathbf{1}_N =0 ,\quad
D + \frac{1}{f(\vtheta )}(\sigma +M) =0.
\label{eq:saddle_Higgs}
\end{\eq}
With explicit indices, the third equation is
\begin{\eq}
\sigma_{ij} \phi_{jA} +\phi_{iB} M_{BA} =0, 
\end{\eq}
where $i,j$ are the gauge indices and $A,B$ are flavor indices.
Let us suppose $N_f \geq N$ and $\phi_{iA}$ is the eigenvector of $M$ with eigenvalues $m_1 ,\cdots ,m_{N_f}$.
Then this equation says that
$\phi_{iA}$ is also the eigenvector of $\sigma$ with eigenvalues $-m_{l_1} ,\cdots ,-m_{l_N}$,
where $(l_1 ,\cdots , l_N )$ is a set of $N$ integers in $( 1,\cdots ,N_f )$.
Thus the localized configuration is labeled by $(l_1 ,\cdots , l_N )$.
Then  up to gauge and flavor rotation, we find 
\begin{\eq}
\sigma_i = -m_{l_i} ,\quad \phi_{iA}= \sqrt{\chi} \delta_{l_i A} .
\label{rootHiggs}
\end{\eq}

\subsubsection*{One-loop determinant}
Let us compute the one-loop determinant around the saddle point \eqref{eq:saddle_Higgs}.
The analysis 
in the Coulomb branch localization
is quite useful also for this purpose.
Now we have three types of fields: vector multiplet, chiral multiplets with vanishing VEV and nonzero VEV.
First two types of one-loop determinants are obtained by just taking $\sigma_i = -m_{l_i}$ in the result of Coulomb branch 
since the new deformation term $V_H $ does not have any derivative terms.
The last one-loop determinant coming from the chiral multiplet with non-vanishing VEV is nontrivial for arbitrary value of $\chi$.
However, since we can ignore the derivative terms in the limit $\chi\rightarrow \pm\infty$,
the one-loop determinant should be unity in this limit.
Thus we conclude that the one-loop determinants on the Higgs branch are
\begin{\eqa}
Z_{\rm vec}^{\rm (1-loop )} 
&=& \prod_{i<j} \sinh{\pi b (m_{l_i}-m_{l_j}  )} \sinh{\pi b^{-1} (m_{l_i}-m_{l_j}  )}  ,\NN\\
Z_{\rm chi}^{\rm (1-loop )} 
&=&   \prod_{A\neq \{ l_i\} } \prod_{i=1}^N s_b \left( \frac{iQ}{2}+ m_{l_i}-m_A  \right) .
\end{\eqa}
For general $\Delta$, the analytic continuation of 
the real
masses leads us to
\begin{\eqa}
Z_{\rm vec}^{\rm (1-loop )} 
&=& \prod_{i<j} \sinh{\pi b (m_{l_i}-m_{l_j}  )} \sinh{\pi b^{-1} (m_{l_i}-m_{l_j}  )}  ,\NN\\
Z_{\rm chi}^{\rm (1-loop )} 
&=&   \prod_{A\neq \{ l_i\} } \prod_{i=1}^N s_b \left( \frac{iQ}{2}(1-\Delta ) + m_{l_i}-m_A  \right) .
\end{\eqa}

\subsection{At north and south poles}
At the north and south poles, we allow singular configurations.
At the north pole ($\vtheta =0$), we have
\begin{\eqa}
&& F_{12}=0,\quad F_{31} =0,\quad
   F_{23}  -\frac{1}{2}(\phi \phi^\dag -\chi \mathbf{1}_N ) =0,\quad  D_\mu \sigma =0,\quad (\sigma +M) \phi =0, \NN\\
&& D + \frac{1}{b}\sigma +\frac{i}{2}(\phi \phi^\dag -\chi \mathbf{1}_N) =0, \quad
D_2 \phi +iD_3 \phi =0,\quad D_1 \phi  =0,\quad  F=0 ,
\end{\eqa}
corresponding to the vortex equation 
while we have at the south pole ($\vtheta =\pi /2$)
\begin{\eqa}
&& F_{12}=0,\quad  F_{23}  =0,\quad
   -F_{31} -\frac{1}{2}(\phi \phi^\dag -\chi \mathbf{1}_N ) =0,\quad  D_\mu \sigma =0,\quad  (\sigma +M) \phi =0, \NN\\
&&D + \frac{1}{b^{-1}}\sigma +\frac{i}{2}(\phi \phi^\dag -\chi\mathbf{1}_N ) =0, \quad
 D_1 \phi  +iD_3 \phi =0,\quad D_2 \phi =0,\quad  F=0,
\end{\eqa}
corresponding to the anti-vortex equation.
In the $\chi \rightarrow\pm \infty$ limit, the size of the (anti-)vortex becomes point-like.
Thus we find (anti-)vortex partition function on $\mathbb{R}^2 \times S_\beta^1$ 
with the radius\footnote{
Note that the fiber direction is $\varphi_2 =\psi +\phi\, (\varphi_1 =\psi -\phi)$ on the North (South) pole. 
} $\beta =2\pi b^{-1} \,(2\pi b)$ and omega deformation parameter $\epsilon = ib^{-1} \,(ib)$ on the North (South) pole as read from \eqref{Qsq}.

Finally let us add anti-fundamental chiral multiplets and one adjoint chiral multiplet.
One can show that
fundamental, anti-fundamental and adjoint scalars cannot have VEV simultaneously.
This reflects a fact that 
anti-fundamental and adjoint scalars can contribute only to the Fermionic moduli \cite{Fujimori:2012ab}.
Therefore away from the north and south poles,
the one-loop determinant for each anti-fundamental or 
adjoint chiral multiplet 
with $\Delta =0$ 
is just\footnote{
One of the easiest ways to find an expression for general R-charge $\Delta$
is to turn on axial mass $\mu$ and 
perform the analytic continuation: $\mu\rightarrow \mu +iQ\Delta /2$.
}
\begin{\eqa}
Z_{\rm chi}^{\rm (1-loop )} 
&=&  \left.  \prod_{w\in R} s_b \left( \frac{iQ}{2} -w(\hat{\sigma })  \right) \right|_{\sigma_i =-m_{l_i}}  .
\end{\eqa}
As we will see in the next section, 
these multiplets nontrivially contribute to vortex partition functions.

To summarize,
as a result of our Higgs branch localization,
it turns out that
the whole partition function takes the following factorized form
\begin{\eq}
Z = \sum_{ \rm Higgs\ branch} Z_{\rm classical}  Z_{\rm vec}^{\rm (1-loop )}  Z_{\rm chi}^{\rm (1-loop )} 
      Z_{\rm V}^{(\{ l_i\})}\bar{Z}_{\rm V}^{(\{ l_i \} )} ,
\label{eq:summary}
\end{\eq}
where $Z_{\rm classical}$ is contributions from the CS and FI-terms at Higgs branch solution.
Since we have already obtained the explicit forms of $Z_{\rm classical}$, $Z_{\rm vec}^{\rm (1-loop )}$ and $Z_{\rm chi}^{\rm (1-loop )}$,
our remaining task is to compute the vortex partition functions $Z_{\rm V}^{(\{ l_i\})}$ and $\bar{Z}_{\rm V}^{(\{ l_i \} )}$.
In next section, we will explicitly calculate $Z_{\rm V}^{(\{ l_i\})}$ and $\bar{Z}_{\rm V}^{(\{ l_i \} )}$
by performing localization of world line theories of the vortices.

\subsubsection*{Remarks}
It was discussed in \cite{Pasquetti:2011fj} that
the Coulomb branch localization formula for $U(1)$ gauge theory with $N_f$ fundamental and anti-fundamental chiral multiplets
exhibits the factorization only for physical integer CS level.
A counterpart of this in the Higgs branch localization is that
localization procedure itself is failed for unphysical CS level due to breaking of the gauge symmetry.
Hence we conclude that 
the factorization formula \eqref{eq:summary} is valid when effective CS level is integer.

We have seen that 
the new deformation term $\mathcal{L}_H$ in \eqref{LH} leads us to
the (anti-)vortex configuration at the north (south) pole. 
If we replace the function $h$ by $-h$,
then we find the vortex at the south pole and anti-vortex at the north pole.
Note that the whole partition function is invariant under this replacement 
since $h$ appears only in the $Q$-exact term.

\section{Vortex partition function and localization}
In the previous section, we have seen that 
the new deformation term \eqref{LH} leads us 
to the (anti-)vortex at the north(south) pole on $\mathbb{R}^2 \times \mathbb{S}^1$ as the localized configuration.
In this section we therefore compute the vortex partition function in a class of 3d $\mathcal{N}=2$ theories.

In \cite{Hanany:2003hp}, the authors have discussed that
the vortex moduli space of non-Abelian gauge theory with eight supersymmetries 
are determined by a brane construction and 
the vortex world line theory preserves 2d $\mathcal{N}=(2,2)$ type supersymmetry. 
Meanwhile it has turned out that the  moduli space of non-Abelian gauge theory is 
constructed in a purely field theoretic manner \cite{Eto:2005yh}.  
Furthermore the authors in \cite{Edalati:2007vk} have analyzed
the half BPS vortices for gauge theory with four supercharges and 
found that the vortices preserve the 2d $\mathcal{N}=(0,2)$ type supersymmetry.
The half BPS vortex world line theory for 3d $\mathcal{N}=2$ SUSY gauge theory  preserves two supercharges,
which are the dimensional reduction of $\mathcal{N}=(0,2)$ supersymmetry from two dimensions to one dimension.
Thus
we can evaluate  partition function of this quantum mechanics called K-theoretic vortex partition function
in a similar manner as the 5d instanton partition function \cite{Nekrasov:2002qd,Nekrasov:2003rj}.

\subsection{Vortex partition function of
3d $\mathcal{N}=2$ theory with fundamental and anti-fundamental
chiral multiplets}
We take the vortex number  as $k$ for gauge field strength along $\R^2 \subset \R^2 \times \mathbb{S}^1 $: 
\bea
k=\frac{1}{2 \pi} \int_{\R^2} \mathrm{Tr}_N F_A. 
\eea
Then 
the vortex quantum mechanics for 3d $\mathcal{N}=2$ supersymmetric gauge theory with $N_f$ fundamental and  $\bar{N}_f$ 
anti-fundamental chiral multiplets  consists of the following multiplets:
\begin{itemize}
\item $U(k)$ vector multiplet: $(A_t, \varphi, D, \lambda^{+}, \bar{\lambda}^{+})$
\item an adjoint chiral multiplet: $(B,\bar{B}, \psi^{-} , \bar{\psi}^{-})$
\item fundamental chiral multiplets with $N_c$-flavors: 
$(I^i, \bar{I}^i, \psi_{I}^{i\,-} , \bar{\psi}_{I}^{i\,-}),\quad i=1,\cdots,N_c$
\item anti-fundamental chiral multiplets with $N_f -N_c$-flavors: \\
$(J^j, \bar{J}^j, \psi_{J}^{j\,-} , \bar{\psi}_{J}^{j\,-}),\quad j=N+1,\cdots,N_f$
\item fundamental Fermi multiplets with $\bar{N}_f$-flavors: 
$(\psi^{p +}, \bar{\psi}^{ p +}, F^{p}, F^{p}),\quad p=1,\cdots, \bar{N}_f $
\end{itemize}
The supersymmetric transformations for these fields are given by\footnote{
We have performed dimensional reduction along the 2-direction
in the manner of \cite{Benini:2013nda}.
}:
\begin{\eq}
\begin{split}
&\delta A_{\omega}=-\frac{i}{2} (\bar{\epsilon}^+ \lambda^+ + \bar{\lambda}^+ \epsilon^+ ), \quad \delta A_{\bar{\omega}} =0, \\
&\delta \bar{\lambda}^+ =\bar{\epsilon}^+(D-iF_{12}),\quad \delta(D-iF_{12})=-2i \epsilon^+ D_{\bar{\omega}}\bar{\lambda}^+, \\
&\delta \lambda^+ =\epsilon^+ (-D-iF_{12}),\quad \delta(-D-iF_{12})=-2i \bar{\epsilon}^+ D_{\bar{\omega}}\lambda^+ , \label{vec-tr}
\end{split}
\end{\eq}
\begin{\eq}
\begin{split}
&\delta B=-\bar{\epsilon}^+\lambda^-,\quad \delta \lambda^- =i \epsilon^+ ( 2 D_{\bar{\omega}}B - \vep B),\\
&\delta \bar{B}= \epsilon^+ \bar{\lambda}^- ,\quad \delta\bar{\lambda}^- =-i \bar{\epsilon}^+ (2 D_{\bar{\omega}} \bar{B}+ \vep \bar{B}) , \label{adj-tr}
\end{split}
\end{\eq}
\begin{\eq}
\begin{split}
&\delta I =-\bar{\epsilon}^+ \psi^{ -}_I, \quad \delta\psi^{ -}_I=i\epsilon^+ (2  D_{\bar{\omega}}I- I m),\\
&\delta \bar{I}=-\epsilon^+ \bar{\psi}^{ -}_I,\quad \delta \bar{\psi}^-_I =i\bar{\epsilon}^+ (2 D_{\bar{\omega}}\bar{I} +m \bar{I} ) , \label{I-tr}
\end{split}
\end{\eq} 
\begin{\eq}
\begin{split}
&\delta J =-\bar{\epsilon}^+ \psi^{ -}_J, \quad \delta\psi^{ -}_J=i\epsilon^+ (2  D_{\bar{\omega}}J + J\tilde{m}),\\
&\delta \bar{J}=-\epsilon^+ \bar{\psi}^{ -}_J,\quad \delta \bar{\psi}^-_J =i\bar{\epsilon}^+ (2 D_{\bar{\omega}}\bar{J} - \tilde{m}\bar{J}) , \label{J-tr}
\end{split}
\end{\eq} 
\begin{\eq}
\begin{split}
&\delta \psi^+ =-\bar{\epsilon}^+ E+\bar{\epsilon}^+ F, \quad \delta  F=i\epsilon^+ (-2  D_{\bar{\omega}}\psi^+ + \tilde{M} \psi^+ + \psi^-_E),\\
&\delta \bar{\psi}^+ =-\bar{\epsilon}^+ \bar{E} +\bar{\epsilon}^+ \bar{F}, \quad \delta  
\bar{F}=i\epsilon^+ (-2  D_{\bar{\omega}} \bar{\psi}^+ -  \psi^+ \tilde{M} + \bar{\psi}^-_E), \label{Fermi-tr}
\end{split}
\end{\eq} 
where $\vep$ is the $\Omega$-background parameter 
for regularizing the flat direction of the adjoint fields, 
$m (\tilde{m})$ and $\tilde{M}$ are the twisted masses of 
the (anti-)chiral multiplets and Fermi multiplet, respectively. 
Here we have omitted flavor suffix. 
We also define 
$A_{\omega}:=\frac{1}{2} (A_1-iA_2), A_{\bar{\omega}}:=\frac{1}{2} (A_1+iA_2)$ and $D_{\omega}=\frac{1}{2} (D_1-iD_2), D_{\bar{\omega}}=\frac{1}{2} (D_1+iD_2)$
with $A_1:=A_t, A_2=\varphi$ and $\partial_2 \Phi =0$. 
Roughly speaking, 
the mass parameters $m,\tilde{m}$ and $\tilde{M}$ in the 3d language are as follows.
\begin{itemize}
\item The $N$ twisted masses $m$: 
the real masses $(m_{l_1}, \cdots, m_{l_N})$ of the 3d fundamental chiral multiplet 
satisfying (\ref{rootHiggs}).
For simplicity, we take $(l_1, \cdots,l_N)=(1,\cdots, N)$ in this section. 

\item The $(N_f -N_c )$ twisted masses $\tilde{m}$: 
the real masses for the 3d fundamental chiral multiplets 
which are not $ (m_{l_1}, \cdots, m_{l_N}) $.

\item The $\bar{N}_f$ twisted masses $\tilde{M}$:
the real masses for the 3d anti-fundamental chiral multiplets.
\end{itemize}

We set $\epsilon^+=1, \bar{\epsilon}^+=1$ in the rest of this section.   
Then the Lagrangian of the vortex quantum mechanics is written in the following $Q$-exact manner:
\begin{eqnarray}
\label{Lagvec}
\mathcal{L}_{\text{vec}}&=&\frac{1}{2} \delta \mathrm{Tr}_k    \bar{\lambda}^+ ( D +iF_{12}), \\
\mathcal{L}_{B}&=&\delta \mathrm{Tr}_k  \left(  i \bar{B}  (2 D_{\omega}+\varepsilon) \lambda^-  -i \bar{B}[\lambda^{+}, B] \right), \\
\mathcal{L}_{I}&=&\delta  \left(  i \bar{I} (2D_{\omega}  +m) {\psi}^{-}_I  -i  \bar{I} \lambda^{+} I  \right), \\
\label{LagJ}
\mathcal{L}_{J}&=&\delta  \left(   i J (2D_{\omega} -\tilde{m} ) \bar{\psi}^-_J  -i J \lambda^{+} \bar{J}  \right), \\
\mathcal{L}_{\text{Fermi}}&=&\frac{1}{2} \delta \left( \overline{\delta \psi^+} \cdot \psi^+ + \psi^+  \delta \overline{\bar{\psi}^+} \right).
\end{eqnarray}
The Fayet-Iliopoulos term is also written in $Q$-exact form as
\begin{eqnarray}
\mathcal{L}_{\text{FI}}=-\frac{i\chi}{2} \delta \left( \bar{\lambda}^+ - {\lambda}^+ \right).
\end{eqnarray}
Note that the Chern-Simons term is not $Q$-exact but the $Q$-closed \cite{Kim:2012uz} as
\begin{eqnarray}
\mathcal{L}_{\text{CS}}=2 i \kappa \mathrm{Tr}_k A_{\bar{\omega}}.
\end{eqnarray}
Here $\kappa$ 
corresponds to
the bare Chern-Simons level in the three dimensions.

When we set all the mass parameters to zero, the D-term equation of the vortex quantum mechanics gives the $k$-vortex moduli space:
\begin{eqnarray}
\mathcal{M}^k_{N, N_f}=\Bigl\{ (B,I,J) \Bigl| \, [B, B^{\dagger}]+I \bar{I}-\bar{J} J =\chi 1_{k} \Bigr\}/ U(k). 
\label{Khalerquotient}
\end{eqnarray}
Here $\chi$ is a positive constant. The partition function of the vortex world line is defined as
\begin{eqnarray}
Z^{k}_{\text{V}}=\int \mathcal{D} \Psi \exp \left( \int^{\beta}_{0} dt  (\mathcal{L}_{CS} - t \delta V) \right),
\label{QMpartition}
\end{eqnarray} 
with
\begin{eqnarray}
\delta V= (\mathcal{L}_{\text{vec}}+\mathcal{L}_B+\mathcal{L}_I+ \mathcal{L}_J +\mathcal{L}_{\text{Fermi}} +\mathcal{L}_{\text{FI}}).
\end{eqnarray}
Here $\Psi$ denotes the collection of the fields in the vortex quantum mechanics.
The $\beta$ is the length of compactified circle  of the world line and  identified with the length of  the circle fiber on the points supporting 
point like (anti-)vortex in the 3d ellipsoid.

Let us evaluate the K-theoretic vortex partition function.
First we drop the CS term and we will add this later.
Since the action for this case is written as the $Q$-exact form, 
we can perform the path integral exactly via localization.
Taking $t \to \infty$, the saddle points for bosonic fields are given by zeros of supersymmetric variation (\ref{adj-tr}), (\ref{I-tr}) and (\ref{J-tr}) as
\begin{eqnarray}
 2 D_{\bar{\omega}}B - \vep B=0,   \quad 2 D_{\bar{\omega}}I-m I =0,   \quad 2 D_{\bar{\omega}}J + \tilde{m}J=0.
\label{fixed1}
\end{eqnarray} 
If we take the gauge fixing condition as $\partial_t A_{\bar{\omega}}=0$,
then  (\ref{fixed1}) reduces to the constant matrix valued equations, namely
\begin{eqnarray}
[2 i A_{\bar{\omega}}, B] - \vep B=0,  \quad 2 iA_{\bar{\omega}}   I- I m,   \quad -2 i  J A_{\bar{\omega}} + \tilde{m}J=0.
\label{fixed2}
\end{eqnarray} 
These  equations are the fixed points equation for the vortex moduli space under the equivariant rotation with respect to $U(1)^{N_{f}-1}_{m} \times U(1)_{\varepsilon}$ \cite{Shadchin:2006yz,Yoshida:2011au,Bonelli:2011fq,Fujimori:2012ab}.
 By taking the diagonal gauge for constant mode for $A_{\bar{\omega}}$, the solutions are given by
\begin{eqnarray}
2 iA_{\bar{\omega}, (l,i)}=m_{i} +(l-1) \varepsilon, \quad i=1 \cdots N, \quad l=1, \cdots, k_{i}. 
\end{eqnarray}
The fixed points are classified by $N$-tuple non-negative integers $(k_1, \cdots, k_N)$ with $\sum_{i=1}^{N} k_i=k$,
where
$k_i$ is vorticity for $i$-th diagonal   $U(1) \subset U(N)$. 
By applying the localization formula \cite{Bruzzo:2002xf}, 
the one-loop determinant around the fixed point labeled by $(k_1, \cdots, k_N)$ is given by 
\begin{eqnarray}
&&Z^{(k_{1}, \cdots, k_N)}_{\text{V}}=
\frac{\prod_{i, j=1}^{N} \prod_{l=1}^{k_i} \prod_{\tilde{l}=1}^{k_j}\Det (\partial_t  +2 i A_{\bar{\omega}, (l,i)}-2 i A_{\bar{\omega}(\tilde{l},j)})}
{\prod_{i, j=1}^{N} \prod_{l=1}^{k_i} \prod_{\tilde{l}=1}^{k_j} \Det ( \partial_t +2 i A_{\bar{\omega}, (l,i)}-2 i A_{\bar{\omega}, (\tilde{l},j)} -\vep )} \nonumber \\
&&  \times \frac{\prod_{i=1 }^{N} \prod_{p=1}^{\bar{N}_f} \prod_{l=1}^{k_{i}} \Det (\partial_t -2 i A_{\bar{\omega}, (l,i)} -M_{p})}
{\prod_{i', j'=1}^{N} \prod_{l'=1}^{k_{i'}} \Det (\partial_t +2 i A_{\bar{\omega}, (l',i')} +m_{j'})
\prod_{i'=1 }^{N} \prod_{j=N+1}^{N_f} \prod_{l'=1}^{k_{i'}} \Det (\partial_t -2 i A_{\bar{\omega}, (l',i')} - \tilde{m}_{j}) }. \nonumber \\
\label{localization1}
\end{eqnarray}
Here the denominator, the numerators in the first and second lines
come from the one-loop determinants of (\ref{Lagvec})-(\ref{LagJ}),
ghost and Fermi multiplet, respectively.
The functional determinant on the circle with radius $\beta$ is evaluated as
\begin{eqnarray}
\mathrm{Det} (\partial_t+ a)=2 \sinh \frac{\beta a}{2}.
\end{eqnarray}
Then, (\ref{localization1}) becomes
\begin{eqnarray}
&&Z^{(k_{1}, \cdots, k_N)}_{\text{V}}=
\frac{\prod_{i, j=1}^{N} \prod_{l=1}^{k_i} \prod_{\tilde{l}=1}^{k_j} 2 \sinh \frac{\beta}{2} (m_{j, i} +(l-\tilde{l}) \vep )}
{\prod_{i, j=1}^{N} \prod_{l=1}^{k_i} \prod_{\tilde{l}=1}^{k_j} 2 \sinh \frac{\beta}{2} ( m_{j, i} +(l-\tilde{l}- 1) \vep   )} \nonumber \\
&&\times \frac{\prod_{i=1 }^{N} \prod_{j=1 }^{\bar{N}_f} \prod_{l=1}^{k_{i}} 2 \sinh \frac{\beta}{2} ( m_{j, i} +(l-1) \vep )}
{\prod_{i, j=1}^{N} \prod_{l=1}^{k_{i}} \sinh \frac{\beta}{2} (m_{j, i} +(l-1) \vep  )
\prod_{i=1 }^{N} \prod_{j=N+1 }^{N_f} \prod_{l=1}^{k_{i}} 2 \sinh \frac{\beta}{2} ( m_{i, j} -(l-1) \vep ) } \nonumber \\
&&=
\frac{\prod_{i=1 }^{N} \prod_{j=1 }^{\bar{N}_f} \prod_{l=1}^{k_{i}} 2 \sinh \frac{\beta}{2} ( m_{j, i} +(l-1) \vep )}
{\prod_{i, j=1}^{N} \prod_{l=1}^{k_i}  2 \sinh \frac{\beta}{2} ( m_{j, i} +(l- 1-k_j) \vep   )
 \prod_{i=1 }^{N} \prod_{j=N+1 }^{N_f} \prod_{l=1}^{k_{i}} 2 \sinh \frac{\beta}{2} ( m_{i, j} -(l-1) \vep ) }. \nonumber \\
\label{localization2}
\end{eqnarray}
Here we define $m_{i j}$ as
\begin{eqnarray}
m_{j, i}=\left \{
\begin{array}{rl}
&m_i- m_j, (i,j \in \{1, \cdots N \}) \\
&m_i- \tilde{m}_j, (i \in \{1, \cdots N \}, j \in \{N+1, \cdots N_f \}) \\ 
&m_i- \tilde{M}_j, (i \in \{1, \cdots N \}, j \in \{1, \cdots \bar{N}_f \}).
\end{array}
\right.
\end{eqnarray}

Next we consider the Chern-Simons term contribution.
The CS term at the fixed point labeled by $(k_1, \cdots, k_N)$ is evaluated as
\begin{eqnarray}
e^{ 2i  \kappa \int \mathrm{Tr} A_{\bar{\omega}}} \Big|_{\text{fixed point}} 
=e^{\left(i \kappa \beta \sum_{i=1}^{N} \sum_{l=1}^{k_i} 2 A_{\bar{\omega}, (l,i) } \right)}
=e^{ \left( \beta \kappa \sum_{i=1}^{N} \left[ k_{i} m_{i} + \vep \frac{k_{i}(k_{i}-1)}{2} \right] \right)}. 
\end{eqnarray}
Therefore, up to over all sign, the K-theoretic vortex partition function for 3d $\mathcal{N}=2$ $U(N)$ Chern-Simons-matter theory with
$N_f$-fundamental and $\bar{N}_f$ anti-fundamental chiral multiplets is given by
\begin{eqnarray}
Z_{\text{V}}&=&\sum_{\{ k_i \}} \left(\prod_{i=1}^{N} z^{k_i}_i \right) 
e^{ \left( \beta \kappa \sum_{i=1}^{N} \left[ k_{i} m_{i} + \vep \frac{k_{i}(k_{i}-1)}{2} \right] \right)} \nonumber \\
&& \frac{\prod_{i=1 }^{N} \prod_{j=1 }^{\bar{N}_f} \prod_{l=1}^{k_{i}} 2\sinh \frac{\beta}{2} ( m_{j, i} +(l-1) \vep )} 
{\prod_{i, j=1}^{N} \prod_{l=1}^{k_i}  2 \sinh \frac{\beta}{2} ( m_{j, i} +(l- 1-k_j) \vep   )
 \prod_{i=1 }^{N} \prod_{j=N+1 }^{N_f} \prod_{l=1}^{k_{i}} 2 \sinh \frac{\beta}{2} ( m_{j, i} +(l-1) \vep ) }. \NN\\
\end{eqnarray}
Here we denoted the 3d complexified FI-parameters as $z_i$.  

\subsubsection*{Remarks}
In this section we have computed 
the K-theoretic vortex partition function by using the vortex quantum mechanics.
One might expect that
we can compute the vortex partition function
also in terms of equivariant character.
However, there is an important difference 
between the results obtained by the vortex quantum mechanics and equivariant character.
Naively, 
the K-theoretic vortex partition function can be obtained 
also
from the equivariant character $ \sum_{i} \pm e^{ \omega_{i, p}}$
by  the replacement $\sum_{i } \pm e^{ \omega_{i,p}} \to \sum_{p} \prod_{i} (1-e^{\omega_{i, p}})^{\mp1}$ as in the case of K-theoretic instanton partition function.
Here $p$ denotes a fixed point under equivariant $U(1)^{N_f-1}_{m} \times U (1)_{\vep}  $ action  and  $\omega_{i,p}$ denotes the equivariant weight at the point $p$.  
If we consider $\bar{N}_f=0$ for simplicity, then 
the equivariant character of tangent space $T\mathcal{M}^{k}_{N, N_f}$ of the vortex moduli space 
at the fixed point labeled by $(k_1, \cdots, k_N)$ is 
\bea
\mathrm{ch} (T \mathcal{M}^{k}_{N, N_f})=\sum_{ i', i=1}^{N} e^{-\beta m_{  i, i'}} \sum_{l=1}^{k_i} e^{-\beta (l-1-k_{i'}) \vep} 
+ \sum_{i=1}^{N} \sum_{j=N+1}^{N_f}  \sum_{l=1}^{k_i} e^{-\beta (m_{i, j} -(l-1) \vep)}.
\eea
Then the one-loop determinant  from the equivariant character is given by
\bea
Z^{(k_{1}, \cdots, k_N)}_{\text{equi-ch}}= \frac{1}{\prod_{i, i'=1}^{N} \prod_{l=1}^{k_i} \left(1-e^{-\beta ( m_{ i', i}+(l-1-k_{i'}) \vep} \right) 
 \prod_{i=1}^{N} \prod_{j=N+1}^{N_f} \prod_{l=1}^{k_i} \left(1- e^{-\beta (m_{i, j} -(l-1) \vep)} \right)}. \nonumber \\
\eea
Note that this expression differs from 
the quantum mechanics result \eqref{localization2}.
The precise relation between them is given by
\bea
Z^{(k_{1}, \cdots, k_N)}_{\text{V}} =  
\left( \prod_{i, j=1}^{N} \prod_{l=1}^{k_i}  e^{- \frac{\beta}{2} ( m_{j,i} +(l- 1-k_j) \vep )} \right)
 \left( \prod_{i=1 }^{N} \prod_{j=N+1 }^{N_f} \prod_{l=1}^{k_{i}} e^{-\frac{\beta}{2}( m_{ i, j} -(l-1) \vep ) } \right)
  Z^{(k_{1}, \cdots, k_N)}_{\text{equi-ch}}. 
\eea
We emphasize that the vortex partition function $Z^{(k_{1}, \cdots, k_N)}_{\text{V}}$
obtained from the quantum mechanics matches 
the K-theoretic vortex partition function appearing in the partition function on the 3d ellipsoid.
It is interesting if we can find some physical reasons for this.

In the two dimensional limit $\beta \to 0$, 
the leading behavior of $\beta$ for the both K-theoretic vortex partition functions
is the same as follows:
\begin{eqnarray}
&&\lim_{\beta \to 0} Z^{(k_{1}, \cdots, k_N)}_{\text{V}} \sim
 \frac{\prod_{i=1 }^{N} \prod_{j=1 }^{\bar{N}_f} \prod_{l=1}^{k_{i}}  ( m_{j, i} +(l-1) \vep )}
{\prod_{i, j=1}^{N} \prod_{l=1}^{k_i}   ( m_{j, i} +(l- 1-k_j) \vep   )
 \prod_{i=1 }^{N} \prod_{j=N+1 }^{N_f} \prod_{l=1}^{k_{i}}  ( m_{i, j} -(l-1) \vep ) }. \nonumber \\
\end{eqnarray}
This correctly reproduces the vortex partition function for the $\mathcal{N}=(2,2)$ supersymmetric gauge theory 
with $N_f$ fundamental and $\bar{N}_f$ anti-fundamental chiral multiplets.

\subsection{Vortex partition function of
3d $\mathcal{N}=2^\ast$ SYM with fundamental hyper multiplets}
Next we consider vortex partition function of  real mass deformation of  
3d $\mathcal{N}=4$ $U(N)$ supersymmetric gauge theory,
whose supersymmetry is often referred to as $\mathcal{N}=2^\ast$.
We take the matter multiplets as $N_f$ fundamental hyper multiplets.
The vortex world line theory preserves  four supercharges which are the 
dimensional reduction of  2d $\mathcal{N}=(2,2)$ type SUSY to one dimension. 
The Lagrangian consists of $\mathcal{N}=(2,2)$ $U(k)$ vector multiplet, an adjoint chiral multiplet, whose lowest component is given by $B$,
fundamental chiral multiplets with $N$-flavors, whose lowest component is $I$  
and  anti-fundamental chiral multiplets with $N_f$-flavors whose lowest component is $J$.
We take the real mass parameter for three dimensional $\mathcal{N}=2$ adjoint chiral multiplet as $m^*$. Then
the $\mathcal{N}=(2,2)$ multiplets split to the $\mathcal{N}=(0,2)$ multiplets.

Since
the supersymmetric variations and Lagrangians of these multiplets can be written in similar manner as the previous subsection, 
we do not repeat this here. 
The fixed point condition for this theory is the same as (\ref{fixed2}):
\begin{eqnarray}
&&Z^{(k_{1}, \cdots, k_N)}_{\text{V}}=
\frac{\prod_{i, j=1}^{N} \prod_{l=1}^{k_i}  2 \sinh \frac{\beta}{2} (m^* +m_{j, i} +(l- 1-k_j) \vep   )}
{\prod_{i, j=1}^{N} \prod_{l=1}^{k_i}  2 \sinh \frac{\beta}{2} ( m_{j, i} +(l- 1-k_j) \vep   )} \nonumber \\
&& \quad \quad \quad \quad \quad \times 
\frac{\prod_{i=1 }^{N} \prod_{j=N+1 }^{N_f} \prod_{l=1}^{k_{i}} 2 \sinh \frac{\beta}{2} (m^*+ m_{i, j} -(l-1) \vep )}
{\prod_{i=1 }^{N} \prod_{j=N+1 }^{N_f} \prod_{l=1}^{k_{i}} 2 \sinh \frac{\beta}{2} ( m_{i, j} -(l-1) \vep ) } .
\label{vortexpartition8}
\end{eqnarray}
In the 2d limit $\beta \to 0$, the leading behavior reproduces
the vortex partition function for $\mathcal{N}=(2,2)^*$ supersymmetric gauge theory considered in \cite{Benini:2012ui} as
\begin{eqnarray}
&& \lim_{\beta \to 0} Z^{(k_{1}, \cdots, k_N)}_{\text{V}} \sim
\frac{\prod_{i, j}^{N} \prod_{l=1}^{k_i}   (m^* +m_{j, i} +(l- 1-k_j) \vep   ) }
{\prod_{i, j}^{N} \prod_{l=1}^{k_i}   ( m_{j, i} +(l- 1-k_j) \vep   )} \nonumber \\
&& \qquad \qquad  \qquad \qquad \times 
\frac{\prod_{i=1 }^{N} \prod_{j=N+1 }^{N_f} \prod_{l=1}^{k_{i}}   (m^*+ m_{i, j} -(l-1) \vep )}
{\prod_{i=1 }^{N} \prod_{j=N+1 }^{N_f} \prod_{l=1}^{k_{i}}  ( m_{i, j} -(l-1) \vep ) } .
\label{vortexpartition9}
\end{eqnarray}

When the real mass parameter $m^*$ goes to zero,  
the supersymmetry of the 3d theory enhances to $\mathcal{N}=4$.
Then the one-loop determinant (\ref{vortexpartition8}) becomes
\bea
Z^{(k_{1}, \cdots, k_N)}_{\text{V}} \Big|_{m^*=0}=1.
\eea
The $k$-vortex partition function $Z^k_{\text{V}}$ is given by
\bea
Z^k_{\text{V}} \Big|_{m^*=0}=\sum_{k_1+\cdots +k_N =k} 1.
\eea
This agrees with the number of possible configurations,
where 
$k$ D1-branes ended on the $N$ D3-branes
in the type IIB brane construction of vortices in the 3d $\mathcal{N}=4$ gauge theory \cite{Hanany:2003hp}.

On the other hand when $m^*$ goes to infinity, 
the leading asymptotic behavior becomes 
\bea
&&Z^{(k_{1}, \cdots, k_N)}_{\text{V}} \sim \frac{1}
{\prod_{i, j=1}^{N} \prod_{l=1}^{k_i}  2 \sinh \frac{\beta}{2} ( m_{j, i} +(l- 1-k_j) \vep   )} \nonumber \\
&& \quad \quad \quad  \quad \quad \quad  \quad \times 
\frac{1}
{\prod_{i=1 }^{N} \prod_{j=N+1 }^{N_f} \prod_{l=1}^{k_{i}} 2 \sinh \frac{\beta}{2} ( m_{i, j} -(l-1) \vep ) } .
\eea
This agrees with the one-loop determinant of the vortex partition function 
without the adjoint and anti-fundamental chiral multiplets.
Note that 
factorization in 3d theory with any adjoint matter 
have not been derived from the Coulomb branch localization yet. 
Hence we conjecture that 
the partition function of the mass deformed $\mathcal{N}=4$ SUSY gauge theory 
is factorized to
the product of  vortex partition function (\ref{vortexpartition8}) and its anti-vortex partner.

\section{Supersymmetric Wilson loop}
Let us consider BPS Wilson loop on the ellipsoid.
We define the supersymmetric Wilson loop in the representation $\mathbf{R}$ as
\begin{\eq}
W_{\mathbf{R}} (C) 
= {\rm Tr}_{\mathbf{R}} \mathcal{P} \exp{\left(  \oint_C  d\tau (iA_\mu \dot{x}^\mu +\sigma |\dot{x}|) \right)} , 
\end{\eq}
where $C$ is the contour of the Wilson loop parametrized by $\tau$ and $\dot{x}^\mu =dx^\mu /d\tau$.
In \cite{Tanaka:2012nr}, the author has argued that
the Wilson loop preserves two supercharges when the contour $C$ is
\begin{\eq}
\varphi_2 (\tau ) = b^{-2} \varphi_1 (\tau )+{\rm Const.} ,\quad \vtheta ={\rm Const.}(\neq 0,\pi /2).
\end{\eq}
Note that this contour becomes the closed loop with a torus knot iff $b^2$ is a rational number.
For the Coulomb branch localization, the VEV of Wilson loop is given by
\begin{\eq}
\langle W_{\mathbf{R}} (C)  \rangle
= \langle {\rm Tr}_{\mathbf{R}} U \rangle , \quad {\rm with}\ U ={\rm diag} (e^{2\pi \sigma^{(0)}_1 } , \cdots ,e^{2\pi \sigma^{(0)}_N } ) .
\end{\eq}
For our Higgs branch localization, the contour $C$ cycles around the north and south poles.
Noting
\begin{\eq}
\oint_{\rm around\ pole} A_\mu dx^\mu = \beta n ,
\end{\eq}
we find that the (unnormalized) VEV is the insertion of
\begin{\eq}
 {\rm Tr}_{\mathbf{R}} U  , \quad {\rm with}\ 
U ={\rm diag} (e^{-2\pi m_{l_1} +2\pi ib^{-1}n_1 +2\pi ib\bar{n}_1 } , \cdots ,e^{-2\pi m_{l_N} +2\pi ib^{-1}n_N +2\pi ib\bar{n}_N } ) .
\end{\eq}
to the vortex and anti-vortex partition functions.
We find that the Wilson loop expectation value are symmetric under the interchange 
between the north and south pole values: $(ib, n) \leftrightarrow (ib^{-1},\bar{n})$.
This is consistent with the observation that the Wilson loops act on holomorphic blocks and anti-holomorphic blocks  in the same way \cite{Beem:2012mb}.

\section{Some examples}
In this section we provide some interesting examples for the ellipsoid case. 
As we have argued above, 
we identify the $\mathbb{S}^1$ radius $\beta$ and omega background parameter $\varepsilon$ in the vortex partition functions
with the fiber radius and angular rotation parameter read from \eqref{Qsq} in the 3d set up, respectively.
Therefore throughout all examples,
we take
\begin{\eqa}
&&\beta=2\pi b^{-1}, \quad \vep=i b^{-1},\quad \text{at north pole ($\theta=0$) },\\
&&\beta=2\pi b, \quad ~~~ \vep=i b ,\quad ~~~ \text{at sorth pole ($\theta=\pi$) }.
\end{\eqa}
Furthermore as we will see later,
we have to take  the equivariant masses in the vortex partition function differently from the ones in the 3d $\mathcal{N}=2$ theories as
$m_i \to m_i+\vep /2 \ (i=1,\cdots,N ) $ and $ m_j \to m_i -\vep /2\  ( j=N+1,\cdots , N_f)$. 
Although similar mass shifts have been observed 
in instanton partition function of 4d $\mathcal{N}=2^\ast$ theory on $\mathbb{S}^4$ \cite{Okuda:2010ke},
we have not found its physical interpretation yet. 
It is interesting if we find any physical origin of this shift. 
Also, since $Z_{\text{V}}^{\{l_i\}}$ and $\bar{Z}_{\text{V}}^{\{l_i\}}$ are interchanged under $b\rightarrow b^{-1}$ with each other,
\begin{\eq}
Z_{\text{V}}^{\{l_i\}} = \left. \bar{Z}_{\text{V}}^{\{l_i\}} \right|_{b\rightarrow b^{-1}} ,\quad
\bar{Z}_{\text{V}}^{\{l_i\}} = \left. Z_{\text{V}}^{\{l_i\}} \right|_{b\rightarrow b^{-1}} ,
\end{\eq}
we will explicitly write down only $\bar{Z}_{\text{V}}^{\{l_i\}}$ in all examples.

\subsection{$U(1)$ gauge theories }
\subsubsection{$U(1)$ gauge theory with $2N_f$ fundamental chiral multiplets }
\label{sec:U1vector}
Let us consider $U(1)$ gauge theory with $2N_f$ fundamental chiral multiplets of charge $+1$,
which is called ``chiral theory'' in \cite{Pasquetti:2011fj}. 
Here we take even number fundamental chiral multiplets
since otherwise we have the parity anomaly. 
The total expression of the partition function with $\Delta =0$ is 
\begin{\eqa}
Z 
= \sum_{i=1}^{2N_f}  e^{-2\pi i\zeta m_i} \prod_{A \neq i}^{2N_f} \, s_b \left( \frac{iQ}{2} +E_{Ai} \right) 
            Z_{\rm V}^{(i)} \bar{Z}_{\rm V}^{(i)}  ,
\end{\eqa}
where $E_{ji} =-(m_j -m_i )$ and
\begin{\eqa}
&&\bar{Z}_{\rm V}^{(i)}
=\sum_{n=0}^{\infty}
      \frac{e^{-2\pi \zeta b n } }{\prod_{l=1}^{n} 2 \sinh \pi i b^2(l-1-n)
                                     \prod_{l=1}^{n}\prod_{j \neq i}^{2N_f}2\sinh \pi b(E_{ji}+i l b)} \, .
\label{eq:U1chiralV}
\end{\eqa}
We can check\footnote{
Note that our convention for the mass is related to the one in \cite{Pasquetti:2011fj} by $m_i = -\mu_i /2$.
} that
the above vortex partition functions agree with
the one obtained from explicit evaluation of the Coulomb branch formula \cite{Pasquetti:2011fj}. 
Note that our vortex (anti-vortex) partition function corresponds to the anti-vortex (vortex) partition function called in \cite{Pasquetti:2011fj}.
While the Coulomb branch localization cannot distinguish whether which is the vortex or anti-vortex,
this is manifest in the Higgs branch localization. 
Of course this difference is physically irrelevant 
since the total expression is given by the product of the vortex and anti-vortex partition functions.
Indeed we can also find the vortex (anti-vortex) at the south (north) pole if we took $h$ in \eqref{LH} as $h\rightarrow -h$
as we have noted in the last of sec.~\ref{sec:Higgs_loc}.

In order to obtain a result with general R-charge,
we should replace $m_i$ by $m_i +i\Delta Q/2$:
\begin{\eqa}
Z 
= \sum_{i=1}^{2N_f}  e^{-2\pi i\zeta ( m_i +i\Delta Q/2 )} \prod_{A \neq i}^{2N_f} \, s_b \left( \frac{iQ}{2} +E_{Ai} \right) 
            Z_{\rm V}^{(i)} \bar{Z}_{\rm V}^{(i)}  ,
\end{\eqa}
where $\bar{Z}_{\rm V}^{(i)}$ is the same as \eqref{eq:U1chiralV}.

\subsubsection{$U(1)$ gauge theory with $N_f$ fundamental and anti-fundamental chiral multiplets }
\label{sec:U1chiral}
Next we consider $U(1)$ gauge theory with $N_f$ fundamental of charge $+1$ with mass $\mu_A +m_A $ and
anti-fundamental chiral multiplets of charge $-1$ with mass $\mu_A -m_A $,
which is referred to as ``non-chiral theory'' in \cite{Pasquetti:2011fj}.
The total partition function for $\Delta =0$ is given by
\begin{\eqa}
Z 
= \sum_{i=1}^{N_f} \frac{ e^{-2\pi i\zeta (\mu_i + m_i )} }{s_b\left( -\frac{iQ}{2} +C_{ii} \right) }
 \prod_{A\neq i}^{N_f} \, 
            \frac{s_b \left( \frac{iQ}{2} +D_{Ai} \right)}{s_b\left( -\frac{iQ}{2} +C_{Ai} \right) }
            Z_{\rm V}^{(i)} \bar{Z}_{\rm V}^{(i)}  ,
\end{\eqa}
where $D_{ji}=-(\mu_j +m_j -\mu_i -m_i )$, $C_{ji}=-(\mu_j -m_j -\mu_i -m_i )$
and 
\begin{\eqa}
&&\bar{Z}_{\rm V}^{(i)}
=\sum_{n=0}^{\infty}
\frac{  e^{-2\pi \zeta b n }  \prod_{l=1}^{n} \prod_{j=1 }^{N_f}  2\sinh \pi b ( C_{j i} +(l-1) ib )} 
      {\prod_{l=1}^{n} 2 \sinh \pi i b^{2}(l-1-n)  \prod_{l=1}^{n}\prod_{j \neq i}^{N_f}2\sinh \pi b(D_{ji}+ilb)}\, ,                                  
\end{\eqa}
which agrees with the result of \cite{Pasquetti:2011fj}.

Making the analytic continuation $m_i \rightarrow m_i +iQ\Delta /2$
leads us to the general $\Delta$ formula: 
\begin{\eqa}
Z 
= \sum_{i=1}^{N_f} \frac{ e^{-2\pi i\zeta (\mu_i + m_i +iQ\Delta /2)} }{s_b\left( -\frac{iQ}{2}(1-2\Delta ) +C_{ii} \right) }
 \prod_{A\neq i}^{N_f} \, 
            \frac{s_b \left( \frac{iQ}{2} +D_{Ai} \right)}{s_b\left( -\frac{iQ}{2}(1-2\Delta ) +C_{Ai} \right) }
            Z_{\rm V}^{(i)} \bar{Z}_{\rm V}^{(i)}  ,
\end{\eqa}
where
\begin{\eq}
\bar{Z}_{\rm V}^{(i)}
=\sum_{n=0}^{\infty}
\frac{  e^{-2\pi \zeta b n }  \prod_{l=1}^{n} \prod_{j=1 }^{N_f}  2\sinh \pi b ( C_{j i} +iQ\Delta +(l-1) ib )} 
      {\prod_{l=1}^{n} 2 \sinh \pi i b^{2}(l-1-n)  \prod_{l=1}^{n}\prod_{j \neq i}^{N_f}2\sinh \pi b(D_{ji}+ilb)}\, .                               
\end{\eq}

\subsection{$U(N)$ gauge theories}
\subsubsection{$U(N)$ gauge theory with with $2N_f$ fundamental chiral multiplets}
Next we extend the setup in sec.~\ref{sec:U1chiral} to $U(N)$ gauge group.
The whole part of the partition function is
\begin{\eqa}
Z 
&=& \sum_{(l_1 ,\cdots ,l_N )\subset (1,\cdots ,2N_f ) }e^{ -2\pi i \zeta \sum_i (m_{l_i} +i\Delta Q/2)} 
      \prod_{i<j} \Bigl[  \sinh{(\pi b E_{l_i l_j}  )} \sinh{(\pi b^{-1} E_{l_i l_j})} \Bigr] \NN\\
&&~~~~~~~~~\times \prod_{i=1}^N \prod_{A \neq \{ l_i \}}^{2N_f} s_b \left( \frac{iQ}{2} +E_{A l_i} \right) Z_{\rm V}^{\{l_i\}} \bar{Z}_{\rm V}^{\{l_i\}}  ,
\end{\eqa}
where $\bar{Z}_{\text{V}}^{\{l_i\}}$ is
\begin{\eq}
\bar{Z}_{\text{V}}^{\{l_i\}}  
= \sum_{\{k_i\}} 
\frac{ \prod_{i=1}^{N} e^{-2\pi \zeta b k_i } } 
{\prod_{i, j}^{N} \prod_{l=1}^{k_i}  2 \sinh \pi b ( E_{l_j  l_i} +(l- 1-k_j)ib   )
 \prod_{i=1 }^{N} \prod_{j\neq \{ l_i \} }^{2N_f} \prod_{l=1}^{k_{i}} 2\sinh \pi b ( E_{j l_i} +i l b ) } .
\end{\eq}

\subsubsection{$U(N)$ gauge theory with $N_f$ fundamental and anti-fundamental chiral multiplets}
Let us consider the non-chiral theory with $U(N)$ gauge group studied in sec.~\ref{sec:U1vector} for $U(1)$ case.
Similar consideration leads us to the following total partition function:
\begin{\eqa}
Z 
&=& \sum_{(l_1 ,\cdots ,l_N )\subset (1,\cdots ,N_f ) }e^{ -2\pi i \xi \sum_i ( \mu_{l_i} +m_{l_i}  +i \Delta Q/2)} 
      \prod_{i<j} \Bigl[  \sinh{(\pi b D_{l_i l_j} )} \sinh{(\pi b^{-1}D_{l_i l_j} )} \Bigr] \NN\\
&&~~~~~~~~~~~~~~~~~~   \times \prod_{i=1}^N \frac{ \prod_{A \neq \{ l_i \}}^{N_f} s_b \left( \frac{iQ}{2} +D_{A l_i}  \right)}
     {\prod_{j=1}^{N_f}s_b\left( -\frac{iQ}{2}(1-2\Delta ) +C_{j l_i}  \right)  } Z_{\rm V}^{\{l_i\}} \bar{Z}_{\rm V}^{\{l_i\}}  ,
\end{\eqa}
where 
\begin{\eq}
\bar{Z}_{\text{V}}^{\{l_i\}} 
= \sum_{\{ k_i\}} 
\frac{ \left(\prod_{i=1}^{N} e^{-2\pi \zeta b k_i } \right) \prod_{i=1 }^{N} \prod_{j=1 }^{N_f} \prod_{l=1}^{k_{i}} 2\sinh \pi b ( C_{j, l_i} +iQ\Delta +(l-1) ib )} 
{\prod_{i, j}^{N} \prod_{l=1}^{k_i}  2 \sinh \pi b ( D_{l_j, l_i} +(l- 1-k_j) ib   )
 \prod_{i=1 }^{N} \prod_{j \neq \{ l_i \} }^{N_f} \prod_{l=1}^{k_{i}} 2\sinh \pi b ( D_{j, l_i} +i l b ) }. 
\end{\eq} 

\subsubsection{$U(N)$ gauge theory with $N_f$ fundamental and $\bar{N}_f$ anti-fundamental chiral multiplets}
Here we extend the previous two examples to arbitrary numbers of the fundamental and anti-fundamental chiral multiplets.
Note that $N_f -\bar{N_f}$ must be even number in order to keep the theory gauge invariant\footnote{
If we add the CS term,
the CS level must be integer for even $N_f -\bar{N_f}$ and
half odd for odd $N_f -\bar{N_f}$.
}. 
The whole partition function is 
\begin{\eqa}
Z 
&=& \sum_{(l_1 ,\cdots ,l_N )\subset (1,\cdots ,N_f ) }e^{ -2\pi i \xi \sum_i ( \mu_{l_i} +m_{l_i}  +i \Delta Q/2)} 
      \prod_{i<j} \Bigl[  \sinh{(\pi b D_{l_i l_j} )} \sinh{(\pi b^{-1}D_{l_i l_j} )} \Bigr] \NN\\
&&~~~~~~~~~~~~~~~~~~   \times \prod_{i=1}^N \frac{ \prod_{A \neq \{ l_i \}}^{N_f} s_b \left( \frac{iQ}{2} +D_{A l_i}  \right)}
     {\prod_{j=1}^{\bar{N}_f}s_b\left( -\frac{iQ}{2}(1-2\Delta ) +C_{j l_i}  \right)  } Z_{\rm V}^{\{l_i\}} \bar{Z}_{\rm V}^{\{l_i\}}  ,
\end{\eqa}
with 
\begin{\eq}
\bar{Z}_{\text{V}}^{\{l_i\}} 
= \sum_{\{ k_i\}} 
\frac{ \left(\prod_{i=1}^{N} e^{-2\pi \zeta b k_i } \right) \prod_{i=1 }^{N} \prod_{j=1 }^{\bar{N}_f} \prod_{l=1}^{k_{i}} 2\sinh \pi b ( C_{j, l_i} +iQ\Delta +(l-1) ib )} 
{\prod_{i, j}^{N} \prod_{l=1}^{k_i}  2 \sinh \pi b ( D_{l_j, l_i} +(l- 1-k_j) ib   )
 \prod_{i=1 }^{N} \prod_{j \neq \{ l_i \} }^{N_f} \prod_{l=1}^{k_{i}} 2\sinh \pi b ( D_{j, l_i} +i l b ) }. 
\end{\eq} 
We emphasize that
the factorization of this theory except $N_f =\bar{N}_f$ and $\bar{N}_f =0$ has not been 
obtained from the Coulomb branch localization formula.

\section{Superconformal index}
Let us consider the 3d superconformal index, or equivalently the partition function on $\mathbb{S}^1 \times \mathbb{S}^2$, 
whose Coulomb branch representation has been obtained in \cite{Kim:2009wb,Drukker:2012sr}.
Explicit evaluations of the matrix integrals show that
``chiral theory''  and ``non-chiral theory''
exhibit
the factorization \eqref{eq:factorization}
in terms of the vortex partition functions \cite{Hwang:2012jh} 
as in the 3d ellipsoid case.  

The metric and orthogonal frame on $\mathbb{S}^1 \times \mathbb{S}^2$ are given by
\begin{\eqa}
&&ds^2 =d\tau^2 +R^2(d\theta^2 +\sin^2 \theta d\varphi^2), \quad
e^1=d\tau,\quad e^2=R d\theta,\quad e^3=R \sin\theta d\varphi,
\end{\eqa}
where $\tau \sim \tau +\beta R$. 
Killing spinors on this manifold satisfy
\begin{\eq}
D_\mu \eps =-\frac{1}{2R}\gamma_\mu \gamma_{1}\eps,\quad
D_\mu \beps =\frac{1}{2R}\gamma_\mu \gamma_{1}\beps ,
\end{\eq}
which are solved by
\begin{\eq}
\epsilon=\frac{1}{\sqrt{2}} e^{-\tau /2R} \begin{pmatrix}
-e^{\frac{i}{2}(\theta -\varphi)} \\
e^{\frac{i}{2}(-\theta -\varphi)}
\end{pmatrix}
,\quad \overline{\epsilon}=\frac{1}{\sqrt{2}} e^{\tau /2R}\begin{pmatrix}
e^{\frac{i}{2}(-\theta +\varphi)} \\
e^{\frac{i}{2}(\theta +\varphi)}
\end{pmatrix} .
\end{\eq}  
Note that these solutions satisfy the following twisted boundary condition:
\begin{\eq}
\eps (\tau +\beta R ) = e^{-\frac{\beta}{2}} \eps (\tau ),\quad
\beps (\tau +\beta R) = e^{+\frac{\beta}{2}} \beps (\tau ) .
\end{\eq} 
In order to construct supersymmtric field theory with $\eps$ and $\beps$,
we should impose such twisted boundary conditions also for fields.
Noting quantum numbers of $\eps$ and $\beps$ as
\begin{\eqa}
&& \mathcal{R}(\eps ) = + \eps ,\quad j_3 (\eps ) =-\frac{1}{2}\eps ,\quad F_i (\eps )=0 ,\NN\\
&& \mathcal{R}(\beps ) = - \beps ,\quad j_3 (\beps ) =+\frac{1}{2}\beps ,\quad F_i (\beps )=0 ,
\end{\eqa}
we can understand the boundary conditions as
\begin{\eq}
\eps (\tau +\beta R) = e^{(-\mathcal{R}-j_3 )\beta_1 +j_3 \beta_2 +F_i M_i } \eps (\tau ),\quad
\beps (\tau +\beta R) = e^{(-\mathcal{R}-j_3 )\beta_1 +j_3 \beta_2 +F_i M_i } \beps (\tau ) ,
\end{\eq} 
where $\beta =\beta_1 +\beta_2$.
Thus if we impose these conditions also for fields as
\begin{\eq}
({\rm fields})_{\tau +\beta R} =e^{(-\mathcal{R}-j_3 )\beta_1 +j_3 \beta_2 +F_i M_i } ({\rm fields})_{\tau},
\end{\eq}
then we can construct consistent supersymmetric field theory 
on $\mathbb{S}^1 \times \mathbb{S}^2$.
It is known that the partition function with such boundary conditions becomes
so-called superconformal index:
\begin{\eq}
\mathcal{I} = {\rm Tr}\Bigl[ (-1)^F e^{-\beta_1 (H-\mathcal{R}-j_3 )} e^{-\beta_2 (H+j_3 )} \prod_i e^{-F_i M_i} \Bigr] .
\end{\eq}

Let us compute the index by using the Higgs branch localization\footnote{
In the rest of this section we take $R=1$.
As in the ellipsoid case, we can easily recover R-dependence.
}. 
Here we consider the $\mathcal{N}=2$ theory with $N_f$ fundamental chiral multiplets\footnote{
One can derive similar results on the other theories
as we have explicitly discussed in the ellipsoid case.
}.
For the Coulomb branch localization, 
the authors in \cite{Drukker:2012sr} have taken the deformation terms as (\ref{def-YM}) plus (\ref{def-mat}),
which are the same as the Ellipsoid case up to the difference between Killing spinor equations (see appendix.~\ref{app:SUSY} for detail). 
For our Higgs branch localization,
we further add the following deformation term  
\begin{\eqa}
\mathcal{L}_H= Q \Tr \Big{[} \frac{(\eps^\dagger \lambda-\beps^\dagger \bar{\lambda})h}{2i} \Big{]} .
\end{\eqa}
Again completing the bosonic part of $\mathcal{L}_{\rm YM} +\mathcal{L}_H $ leads us to
\begin{\eqa}
&& \left. \mathcal{L}_{\rm YM} \right|_{\rm Bos.}  +\left. \mathcal{L}_H \right|_{\rm Bos.} \NN\\
&&= {\rm Tr}\Bigl[ \frac{1}{4}( V_1 +\cos{\theta} h )^2 +\frac{1}{4}( V_2 -\sin{\theta} h )^2 +\frac{1}{4}V_3^2 \NN\\ 
&&~~~~~~             +\frac{1}{4}( \bar{V}_1 +\cos{\theta} h )^2 +\frac{1}{4}( \bar{V}_2 +\sin{\theta} h )^2 
                    +\frac{1}{4}\bar{V}_3^2    +\frac{1}{2}(D +ih)^2  \Bigr] ,
\end{\eqa}
where
\begin{\eq}
V_{\hat{\mu}}
=\frac12 \eps_{\hat{\mu} \hat{\nu} \hat{\rho} }F^{\hat{\nu} \hat{\rho} }-D_{\hat{\mu}}\sigma +\delta_{\hat{\mu}1}\sigma , \quad
\bar{V}_{\hat{\mu}}  
= \frac{1}{2} \eps_{\hat{\mu} \hat{\nu} \hat{\rho} }F^{\hat{\nu} \hat{\rho} } + D_{\hat{\mu}} \sigma +\delta_{\hat{\mu} 1}\sigma , 
\end{\eq}
and $\hat{\mu}, \hat{\nu}$ and $\hat{\rho}$ are orthogonal frame indices. 
Combined with $\mathcal{L}_\psi$, the localized configuration is given by
\begin{\eqa}
&& D_1 \phi =0,\quad F=0, \quad
 \sin{\frac{\theta}{2}} D_- \phi  +\cos{\frac{\theta}{2}} (\sigma +M +\Delta )\phi =0,\NN\\
&& \cos{\frac{\theta}{2}} D_+ \phi  +\sin{\frac{\theta}{2}} (\sigma +M -\Delta )\phi =0 ,\\\
&& V_1 +\cos{\theta} h =0,\quad V_2 -\sin{\theta} h=0,\quad V_3 =0,\NN\\
&& \bar{V}_1 +\cos{\theta} h =0,\quad \bar{V}_2 +\sin{\theta} h=0,\quad \bar{V}_3 =0 ,
\end{\eqa}
which includes the Higgs branch solution as in the ellipsoid case.
In particular at the north pole ($\theta =0$), we have
\begin{\eqa}
&& F_{12}=0,\quad F_{31} =0,\quad
   F_{23} +\sigma +h =0,\quad  D_\mu \sigma =0,\quad (\sigma +M) \phi =0, \NN\\
&& D + ih =0, \quad
D_2 \phi -iD_3 \phi =0,\quad D_1 \phi  =0,\quad  F=0 ,
\end{\eqa}
corresponding to the vortex equation,   
while we have at the south pole ($\theta =\pi$)
\begin{\eqa}
&& F_{12}=0,\quad  F_{31}  =0,\quad
   F_{23} +\sigma -h =0,\quad  D_\mu \sigma =0,\quad  (\sigma +M) \phi =0, \NN\\
&&D + ih =0, \quad
 D_2 \phi  +iD_3 \phi =0,\quad D_1 \phi =0,\quad  F=0,
\end{\eqa}
which is the anti-vortex equation. 

We again compute the one-loop determinant except the north and south poles 
by using the index theorem and the result on the Coulomb branch localization.
Noting (see \cite{Drukker:2012sr} for detail)
\begin{\eq}
Q^2  = i\mathcal{L}_v +i(iv^\mu A_\mu  +\sigma \beps\eps ) +i\mathcal{R} 
          +i\beta^{-1} [ (-\mathcal{R} -j_3 ) \beta_1 +j_3 \beta_2 ] ,
\label{Qsqind}
\end{\eq}
with
\begin{\eq}
v = \del_\tau -i \beta^{-1} \del_\varphi ,
\end{\eq}
we find
\begin{\eqa}
&&Z_{\rm chi}^{\rm (1-loop )}=\prod_{A\neq \{ l_j \} }^{N_f} \prod_{j=1}^N \Big{(} 
                               \prod_{r=0}^{\infty}\frac{1-x^{2r +2-\Delta}e^{i\beta (M_{l_j} -M_A ) }}
                                                        {1-x^{2r+\Delta } e^{-i\beta (M_{l_j} -M_A )}} \Big{)},\NN\\
&&Z_{\rm vec}^{\rm (1-loop )}=\prod_{i\neq j} \Bigg{[} 2\sinh \Big{(} \frac{i}{2}\beta (M_{l_i} -M_{l_j})  \Big{)} \Bigg{]} ,
\end{\eqa}
where $x=e^{-\beta_2}$. 
Note that for $\Delta =0$ case, this agrees with the perturbative part of the factorized form found in \cite{Hwang:2012jh}.

At the north and south poles, we have the vortex and anti-vortex, respectively. 
From \eqref{Qsqind}, we identify the equivariant parameter as $\eps  =2\beta_2 \beta^{-1}$ for the both cases. 
Thus we reach to the desired vortex partition function:
\begin{\eqa}
&& Z_{V}^{\{ i_\alpha \}} \left( 0, z , \beta , 2\beta_2 \beta^{-1} , i M_{i_\alpha} +\frac{\eps}{2}  , iM_{j\neq i_\alpha} -\frac{\eps}{2} \right) \NN\\
&=&\sum_{n_1 ,\cdots ,n_{N_c}}   \prod_{\alpha =1}^{N_c} 
    \frac{(-1)^{N_f n_\alpha}  z^{n_\alpha} }
            {\prod_{\beta =1}^{N_c} \prod_{l=1}^{n_\alpha} 2\sinh{ \frac{\beta (iM_{i_\alpha ,i_\beta} +2\beta_2  (l-1-n_\beta ) ) }{2} } } 
    \frac{1}{    \prod_{l=1}^{n_\alpha} \prod_{j=1(\neq \{ i_\alpha \} )}^{2N_f} 2\sinh{\frac{ \beta ( iM_{j,i_\alpha } +2\beta_2 ) }{2} } } . \NN\\
\end{\eqa}
This also matches with the vortex part of the factorized form 
obtained from the Coulomb branch localization formula \cite{Hwang:2012jh}.

\section{Conclusion}
In this paper we have discussed
the factorization of the $\mathcal{N}=2$ partition functions on the three-dimensional ellipsoid and $\mathbb{S}^1 \times \mathbb{S}^2$.
While the previous studies \cite{Pasquetti:2011fj,Taki:2013opa,Hwang:2012jh} have found the factorization
by explicitly evaluating the Coulomb branch matrix integrals,
we have directly derived this by performing the Higgs branch localization and
given the physical interpretations of property of the holomorphic block in the language of the original setup.
It has turned out that
the factorization occurs also for the cases
which have not been found so far.
More concretely,
we have derived the factorization 
for the $\mathcal{N}=2\ U(N)$ theory with arbitrary number of the fundamental and anti-fundamental chiral multiplets and
mass deformation of $\mathcal{N}=4$ SQCD.
We emphasize that this is the first observation of the factorization for theory with adjoint matter.
Furthermore we have also discussed that 
the supersymmetric Wilson loop with the torus knot \cite{Tanaka:2012nr} has also the factorization property
in arbitrary representation.

Some obvious applications of our work are to consider other BPS observables and other 3d spaces.
For example we have not discussed BPS vortex loops, 
whose Coulomb branch representations have been already found in \cite{Drukker:2012sr,Kapustin:2012iw}. 
Along the latter direction, 
there are works of Coulomb branch localization on the bi-axially squashed lens space \cite{Imamura:2012rq} and 
a subspace of the round sphere with Dirichlet boundary condition \cite{Sugishita:2013jca}.
In particular the authors in \cite{Imamura:2013qxa} have argued that
the factorization property is a useful criterion to determine relative phases among contributions from different holonomies.
Therefore it is attractive if we directly derive the factorization by using the Higgs branch localization.
The application to the space with the boundary is more challenging and very interesting.
 
While we have treated single $U(N)$ gauge group with fundamental matters and with anti-fundamental matters, 
 it  is possible  to consider   generalizations   to the theories with matters in more complicated  representations of a gauge group or 
quiver gauge theories.  In these cases, BPS (anti-)vortex equations appeared at north (south) pole  are replaced by 
half (anti-)BPS equations for these theories. Then it is expected that the partition function of the zero mode theory 
for half (anti-)BPS equations describes (anti-)holomorphic parts of factorized partition function in three dimensions. 
It is interesting to study the structure of the moduli space of such BPS equations and localization of partition function of the zero mode theory.

Finally we mention to possible insights to string/M-theory.
As well known it is expected that
$\mathcal{N}=3$ circle quiver Chern-Simons matter (CSM) theory is dual to M-theory on certain background 
as represented by the ABJ(M) theory \cite{Aharony:2008ug}.
Although we have not treated this class of theories here,
it is highly illuminating 
if we can relate the factorization property to recent significant progress 
on partition functions \cite{Drukker:2010nc,Fuji:2011km,Herzog:2010hf,Awata:2012jb} and 
BPS Wilson loops \cite{Marino:2009jd} 
in the $\mathcal{N}=3$ quiver CSM on the round sphere.
Since the vortex partition function, or holomorphic block, can be understood from state counting,
it is very interesting to understand the novel Airy function behavior \cite{Fuji:2011km} from this perspective.
Furthermore it would be worth to investigate connection of geometric engineering of the vortex partition function \cite{Dimofte:2010tz,Bonelli:2011fq}
to the relation \cite{Marino:2009jd,Drukker:2010nc} between the ABJ(M) theory and topological string on local $\mathbb{P}^1 \times \mathbb{P}^1$.
This point of view might be evocative of the Ooguri-Strominger-Vafa relation \cite{Ooguri:2004zv}, \cite{Aganagic:2005dh}:
\begin{\eq}
Z_{\rm BH} = |Z_{\rm top}|^2 .
\end{\eq}

\subsection*{Acknowledgment}
The authors would like to thank Naofumi Hama, Kazuo Hosomichi, Yosuke Imamura, Kazutoshi Ohta, 
Norisuke Sakai, Masato Taki and Seiji Terashima for valuable discussions.
M.H. is grateful to Yukawa institute for theoretical physics
for hospitality, where part of this work was done.
M.H. was supported by Grant-in-Aid for JSPS fellows (No.22-2764).
 
\appendix
\section{Conventions for spinors}
Clifford algebra is given by
\begin{\eq}
\{ \gam_\mu , \gam_\nu \} = 2g_{\mu\nu},
\end{\eq}
The gamma matrix $\gam_\mu$ with the curved space index is 
defined by $\gam_\mu = \gam_a e^a_{\ \mu}$ with the Pauli matrix $\gamma_a$ and vielbein $e^a_{\ \mu}$.
We also define spinor contraction as
\begin{\eq}
\bpsi \psi = \bpsi^\alpha C_{\alpha\beta} \psi^\beta ,\quad
\bpsi \gam^\mu \psi = \bpsi^\alpha C_{\alpha\beta} (\gam^\mu )^\beta_{\ \gam} \psi^\gam ,
\end{\eq}
where\begin{\eq}
C = -i\gam_2 = \begin{pmatrix} 0 & -1 \cr 1 & 0\cr \end{pmatrix}.
\end{\eq}
Then for Grassmann-odd spinors $\psi$ and $\bpsi$, we have
\begin{\eq}
\bpsi\psi = \psi\bpsi ,\quad
\bpsi\gam^\mu \psi = -\psi\gam^\mu \bpsi ,\quad
(\gam^\mu \bpsi )\psi = -\bpsi \gam^\mu \psi .
\end{\eq}
Also we distinguish the above component notation from a matrix notation such that, 
\begin{\eqa}
\bar{\psi}\psi=\bar{\psi}^{\text{T}}C\psi ,\quad 
\bpsi\gam^\mu \psi = \bpsi^{\text{T}}C \gam^\mu \psi . 
\end{\eqa}
In particular noting $\eps =C^{-1}\beps^\ast$, $\beps =C\eps^\ast$ for $\mathbb{S}_{b}^{3}$, we find the relation
\begin{\eq}
\begin{split}
&\epsilon^\dagger \gamma^\mu \lambda =(\epsilon^*)^T \lambda 
     =(C\epsilon^*)^T C\gamma^\mu \lambda = (\bar{\epsilon} \gamma^\mu\lambda),  \\
&\bar{\epsilon}^\dagger \gamma^\mu \bar{\lambda}
      =(C\epsilon^*)^\dagger \gamma^\mu \bar{\lambda}=\epsilon^T C^T \gamma^\mu \bar{\lambda}=-(\epsilon\gamma^\mu \bar{\lambda}),  \\
&\epsilon^{\dagger} \lambda = ( C \epsilon^*)^T C \lambda = (\bar{\epsilon} \lambda ),  \\
&\bar{\epsilon}^\dagger \bar{\lambda}=(C\epsilon^*)^\dagger \bar{\lambda}=\epsilon^T C^T \lambda =-(\epsilon \bar{\lambda}),
\end{split}
\end{\eq}
\begin{\eq}
\begin{split}
&\eps^\dag \eps =1=\beps^\dag \beps ,\quad
\eps^\dag \gam_1 \eps =\cos{\vtheta} = -\beps^\dag \gam_1 \beps ,\quad
\eps^\dag \gam_2 \eps = -\sin{\vtheta} = -\beps^\dag \gam_2 \beps ,\\
& \eps^\dag \gam_3 \eps = 0 = \beps^\dag \gam_3 \beps ,\quad
\eps^\dag \beps =0.
\end{split}
\end{\eq}

\section{Supersymmetry on 3d curved manifold}
\label{app:SUSY}
The supersymmetry transformation is generated by supercharges with the Killing spinors on each geometry satisfying
\begin{\eq}
D_\mu \eps =\gamma_\mu \tilde{\eps},\quad
D_\mu \beps = \gamma_\mu \tilde{\beps}.   
\end{\eq}
For $\mathbb{S}_{b}^{3}$, the covariant derivative is defined by turning on a background $U(1)$ gauge field $V=\frac12(1-\frac{b}{f})d\varphi_1 +\frac12 (1-\frac{b^{-1}}{f})d\varphi_2$ additionally.
These expressions for $\mathbb{S}_{b}^{3}$ and $\mathbb{S}^1 \times \mathbb{S}^2$ are explicitly given by
\begin{\eqa}
&&\tilde{\eps}=\frac{i}{2Rf(\vtheta)}\eps,\qquad \tilde{\beps}=\frac{i}{2Rf(\vtheta)}\beps, \quad \text{for $\mathbb{S}_{b}^{3}$,}\\
&&\tilde{\eps}=-\frac{1}{2R}\gamma_1 \eps,\qquad \tilde{\beps}=\frac{1}{2R}\gamma_1 \beps ,
\quad\quad ~ \text{for $\mathbb{S}^1 \times \mathbb{S}^2$}  .
\end{\eqa} 

\subsection{Vector multiplet}
We introduce supersymmetric transformation for the $\mathcal{N}=2$ vector multiplet as 
\begin{\eq}
\begin{split}
\delta A_\mu   &= \frac{i}{2}( \beps \gamma_\mu \lambda -\blam \gam_\mu \eps ), \\
\delta \sigma  &= \frac{1}{2} (\beps \lambda -\blam\eps ), \\
\delta \lambda &= -\frac{1}{2} \gamma^{\mu\nu} \eps F_{\mu\nu} -D\eps +i\gamma^\mu \eps D_\mu \sigma 
                   +\frac{2i}{3}\sigma \gamma^\mu D_\mu \eps ,\\
\delta \blam   &= -\frac{1}{2} \gamma^{\mu\nu} \beps F_{\mu\nu} +D\beps -i\gamma^\mu \beps D_\mu \sigma
                   -\frac{2i}{3}\sigma \gamma^\mu D_\mu \beps  ,\\
\delta D       &= -\frac{i}{2}\beps \gam^\mu D_\mu \lambda -\frac{i}{2}D_\mu \blam \gam^\mu \eps 
                   +\frac{i}{2}[\beps \lambda ,\sigma ] +\frac{i}{2}[\blam\eps ,\sigma ] 
                   -\frac{i}{6}(D_\mu \beps \gamma^\mu \lambda +\blam \gamma^\mu D_\mu \eps),
\end{split}
\end{\eq}
When we decompose it into $\delta =\delta_\eps +\delta_{\beps}$, 
these commutators generate the following algebra
\begin{\eq}
\begin{split}
&[\delta_\eps , \delta_{\eps^\prime} ] ({\rm Any})=0 ,\quad
[\delta_{\beps} , \delta_{\beps^\prime} ] ({\rm Any})=0 ,\\
&[\delta_\eps , \delta_{\beps} ] A_\mu 
  = iv^\nu \del_\nu A_\mu +i\del_\mu v^\nu A_\nu -D_\mu \Lambda ,\\
&[\delta_\eps , \delta_{\beps} ] \sigma
  = iv^\mu \del_\mu \sigma  +i[\Lambda ,\sigma ] +\rho \sigma  ,\\ 
&[\delta_\eps , \delta_{\beps} ] \lambda
  = iv^\mu \del_\mu \lambda +\frac{i}{4}\Theta_{\mu\nu}\gam^{\mu\nu}\lambda +i[\Lambda ,\lambda ] +\frac32 \rho \lambda +\alpha \lambda +\alpha\lambda ,\\ 
&[\delta_\eps , \delta_{\beps} ] \blam
  = iv^\mu \del_\mu \blam +\frac{i}{4}\Theta_{\mu\nu}\gam^{\mu\nu}\blam  +i[\Lambda ,\blam ] +\frac32 \rho \blam -\alpha \blam,  -\alpha\blam ,\\ 
&[\delta_\eps , \delta_{\beps} ] D
  = iv^\mu \del_\mu D  +i[\Lambda ,D ] +2 \rho D +\mathcal{W} , \label{vec-alg}
\end{split}
\end{\eq}
where
\begin{\eq}
\begin{split}
& v^\mu = \beps \gam^\mu \eps ,\quad
\Theta^{\mu\nu} =D^{[\mu}v^{\nu ]} +v^\lam \omega_\lam^{\mu\nu} ,\\
& \Lambda = v^\mu iA_\mu +\sigma \beps\eps ,\quad
\rho=\frac{i}{3}(\beps \gamma^\mu D_\mu \eps +D_\mu \beps \gamma^\mu \eps), \\
&\alpha = \frac{i}{3} (D_\mu \beps \gam^\mu \eps -\beps \gam^\mu D_\mu \eps ) +v^\mu V_\mu , \quad
\mathcal{W}=\frac13\sigma (\beps \gam^\mu \gam^\nu D_\mu D_\nu \eps-\eps \gam^\mu \gam^\nu D_\mu D_\nu \beps). 
\end{split}
\end{\eq} 
Here $V_\mu$ is the background $U(1)$ gauge field and $\omega_{\lambda}^{\mu\nu}$ is the spin connection. As long as we consider 
$\mathbb{S}_{b}^{3}$ and $\mathbb{S}^1 \times \mathbb{S}^2$, 
one easily finds that $\mathcal{W}$ and $\rho$ vanish.  
Since the algebra generates the translation, Lorentz rotation, R-symmetry rotation and gauge transformation, the algebra closes 
off-shell.

From $\delta =\delta_\eps +\delta_{\beps}$,  
we construct ``fermionic`` supercharge $Q_\alpha$ and $\bar{Q}_\alpha$ as
\begin{\eq}
\delta_\eps =\eps^\alpha Q_\alpha ,\quad \delta_{\beps} =\beps^\alpha \bar{Q}_\alpha. \label{Q-def1}
\end{\eq}
Then we introduce fermionic operators $Q$ as
\begin{\eq}
Q= i( \eps^\alpha Q_\alpha +\beps^\alpha \bar{Q}_\alpha ) \label{Q-def2}
\end{\eq}
with the ``commuting`` spinors $\eps$ and $\beps$. 

\subsection{Chiral multiplet}
Supersymmetry transformation for the $\mathcal{N}=2$ chiral multiplet is given by 
\begin{\eq}
\begin{split}
\delta \phi      &= \beps \psi \\
\delta \bar{\phi}&= \eps \bpsi \\
\delta \psi      &= i\gam^\mu \eps D_\mu \phi +i\eps \sigma \phi
                        +\frac{2\Delta i}{3} \gamma^\mu D_\mu \eps \phi +\beps F \\
\delta \bpsi     &= i\gam^\mu \beps D_\mu \bar{\phi} +i\bar{\phi} \sigma \beps
                       +\frac{2\Delta i}{3} \bar{\phi} \gamma^\mu D_\mu \beps +\bar{F}\eps  \\
\delta F         &= \eps (i\gam^\mu D_\mu \psi -i\sigma \psi -i\lambda \phi ) 
                      +\frac{i}{3}(2\Delta-1)D_\mu \eps \gamma^\mu \psi \\
\delta \bar{F}    &= \beps (i\gam^\mu D_\mu \bar{\psi} -i\bar{\psi}\sigma  +i\bar{\phi}\blam ) 
                     +\frac{i}{3}(2\Delta-1)D_\mu \beps \gamma^\mu \bpsi ,
\end{split}
\end{\eq}
The commutators generate the following algebra 
\begin{\eq}
\begin{split}
&[\delta_\eps , \delta_{\eps^\prime} ] ({\rm Any})=0 ,\quad
[\delta_{\beps} , \delta_{\beps^\prime} ] ({\rm Any})=0 ,\\
&[\delta_\eps , \delta_{\beps} ] \phi 
  = iv^\mu \del_\mu \phi +i\Lambda \phi +\Delta \rho \phi - \Delta \alpha \phi ,\\
&[\delta_\eps , \delta_{\beps} ] \bar{\phi}
  = iv^\mu \del_\mu \bar{\phi} - i \bar{\phi} \Lambda +\Delta \rho \bar{\phi} + \Delta \alpha \bar{\phi}  ,\\ 
&[\delta_\eps , \delta_{\beps} ] \psi
  = iv^\mu \del_\mu \psi +\frac{1}{4}\Theta_{\mu\nu}\gam^{\mu\nu}\psi +i\Lambda \psi +\Big{(}\Delta+\frac12 \Big{)}\rho \psi +(1-\Delta)\alpha\psi ,\\ 
&[\delta_\eps , \delta_{\beps} ] \bar{\psi}
  = iv^\mu \del_\mu \bar{\psi} +\frac{1}{4}\Theta_{\mu\nu}\gam^{\mu\nu}\bar{\psi} -i\bar{\psi}\Lambda+\Big{(}\Delta+\frac12 \Big{)}\rho \bpsi +(\Delta-1)\alpha\bar{\psi} ,\\ 
&[\delta_\eps , \delta_{\beps} ] F
  = iv^\mu \del_\mu F  +i\Lambda F +(\Delta+1)\rho F+(2-\Delta)\alpha F , \\
&[\delta_\eps , \delta_{\beps} ] \bar{F}
  = iv^\mu \del_\mu \bar{F} -i\bar{F}\Lambda+(\Delta+1)\rho \bar{F} +(\Delta -2)\alpha \bar{F},
\end{split}
\end{\eq}
One can easily show that this algebra also closes off-shell.


\providecommand{\href}[2]{#2}\begingroup\raggedright\endgroup

\end{document}